%% file: main.tex
\documentclass[conference]{IEEEtran} 
\IEEEoverridecommandlockouts
\usepackage{cite}
\usepackage{amsmath,amssymb,amsfonts}
\usepackage{algorithm}
\usepackage{algorithmic}
\usepackage{graphicx}
\usepackage{textcomp}
\usepackage{xcolor}
\usepackage[caption=false]{subfig}
\usepackage{float}
\usepackage{paralist}
\usepackage{xspace}
\usepackage{setspace}
\usepackage{lipsum}
\makeatletter
\newcommand*\bigcdot{\mathpalette\bigcdot@{.5}}

\newcommand*\bigcdot@[2]{\mathbin{\vcenter{\hbox{\scalebox{#2}{$\m@th#1\bullet$}}}}}
\makeatother

\usepackage{subfig}
\usepackage{mathrsfs}

\usepackage{siunitx}
\usepackage{url}

\usepackage{tikz}
\usepackage{amsmath}
\usepackage{bm}

\newcommand{\oursystem}{\textit{Anteumbler}}

\def\ie{\textit{i.e.}\xspace}

\def\eg{\textit{e.g.}\xspace}

\begin{document}

\title{Anteumbler: Non-Invasive Antenna Orientation Error Measurement for WiFi APs}

\author{\IEEEauthorblockN{Dawei Yan\textsuperscript{1}, Panlong Yang\textsuperscript{2 *}, Fei Shang\textsuperscript{1}, Nikolaos M. Freris\textsuperscript{1}, Yubo Yan\textsuperscript{1}} 
\textsuperscript{1}University of Science and Technology of China, \textsuperscript{2}Nanjing University of Information Science \& Technology  \\
yandw@mail.ustc.edu.cn, shf\_1998@outlook.com, \{plyang, nfr, yuboyan\}@ustc.edu.cn 
\thanks{\textsuperscript{*} Corresponding author}
}

\maketitle

\begin{abstract}
\input{abstract}
\end{abstract}

\input{section/introduction}

\input{section/preliminary}

\input{section/overview}

\input{section/design}

\input{section/implementation}

\input{section/evaluation}

\input{section/related_work}

\input{section/conclusion}

\input{acknow}
\bibliographystyle{IEEEtran}
\bibliography{reference}

\end{document}

%% file: abstract.tex
The performance of WiFi-based localization systems is affected by the spatial accuracy of WiFi AP. Compared with the imprecision of AP location and antenna separation, the imprecision of AP's or antenna's orientation is more important in real scenarios, including AP rotation and antenna irregular tilt. In this paper, we propose \oursystem \ that non-invasively, accurately and efficiently measures the orientation of each antenna in physical space. Based on the fact that the received power is maximized when a Tx-Rx antenna pair is perfectly aligned, we construct a spatial angle model that can obtain the antennas' orientations without prior knowledge. However, the sampling points of traversing the spatial angle need to cover the entire space. We use the orthogonality of antenna directivity and polarization and adopt an iterative algorithm to reduce the sampling points by hundreds of times, which greatly improves the efficiency. To achieve the required antenna orientation accuracy, we eliminate the influence of propagation distance using a dual plane intersection model and filter out ambient noise. Our real-world experiments with six antenna types, two antenna layouts and two antenna separations show that \oursystem \ achieves median errors below \SI{6}{^\circ} for both elevation and azimuth angles, and is robust to NLoS and dynamic environments. Last but not least, for the reverse localization system, we deploy \oursystem \ over LocAP and reduce the antenna separation error by \SI{10}{mm}, while for the user localization system, we deploy \oursystem \ over SpotFi and reduce the user localization error by more than \SI{1}{m}.

%% file: section/introduction.tex
\section{Introduction} \label{sec:introduction}
Due to its ubiquitous infrastructure, WiFi is developing as a candidate for sensing, such as localization~\cite{mDTrack,ArrayTrack,10.1145/2639108.2639142,spotfi}, health monitoring~\cite{wang2016human,ma2022mobi2sense} and imaging~\cite{huang2014feasibility,pallaprolu2022wiffract}. 
Despite the many innovations of these arts, deploying them to sense indoor environments remains a considerably challenging problem. An important reason for this situation is the need to ensure the credibility of the WiFi infrastructure itself, \ie, the requirement for accurate prior knowledge of WiFi APs' or antennas' locations and orientations in the sensing ambiences. Inaccuracies in this knowledge may introduce computational errors that render the sensing systems ineffective. For example, techniques that use antenna arrays to combat multipath rely on precise antenna separation and orientation~\cite{mDTrack,ArrayTrack,spotfi}, and feature-based sensing methods require antennas to remain relatively consistent during the training and test~\cite{shang2022liqray,gao2021towards}.

\begin{figure}[t]
	\centering
	\subfloat[\label{fig:antenna rotation}]{
		\includegraphics[width=0.81\linewidth]{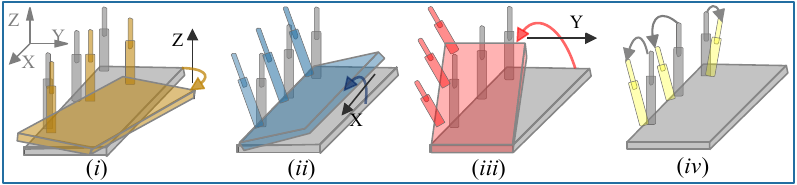}}
		
	\vspace{-0.25cm} 
	\subfloat[\label{fig:AP antenna}]{
		\includegraphics[width=0.36\linewidth]{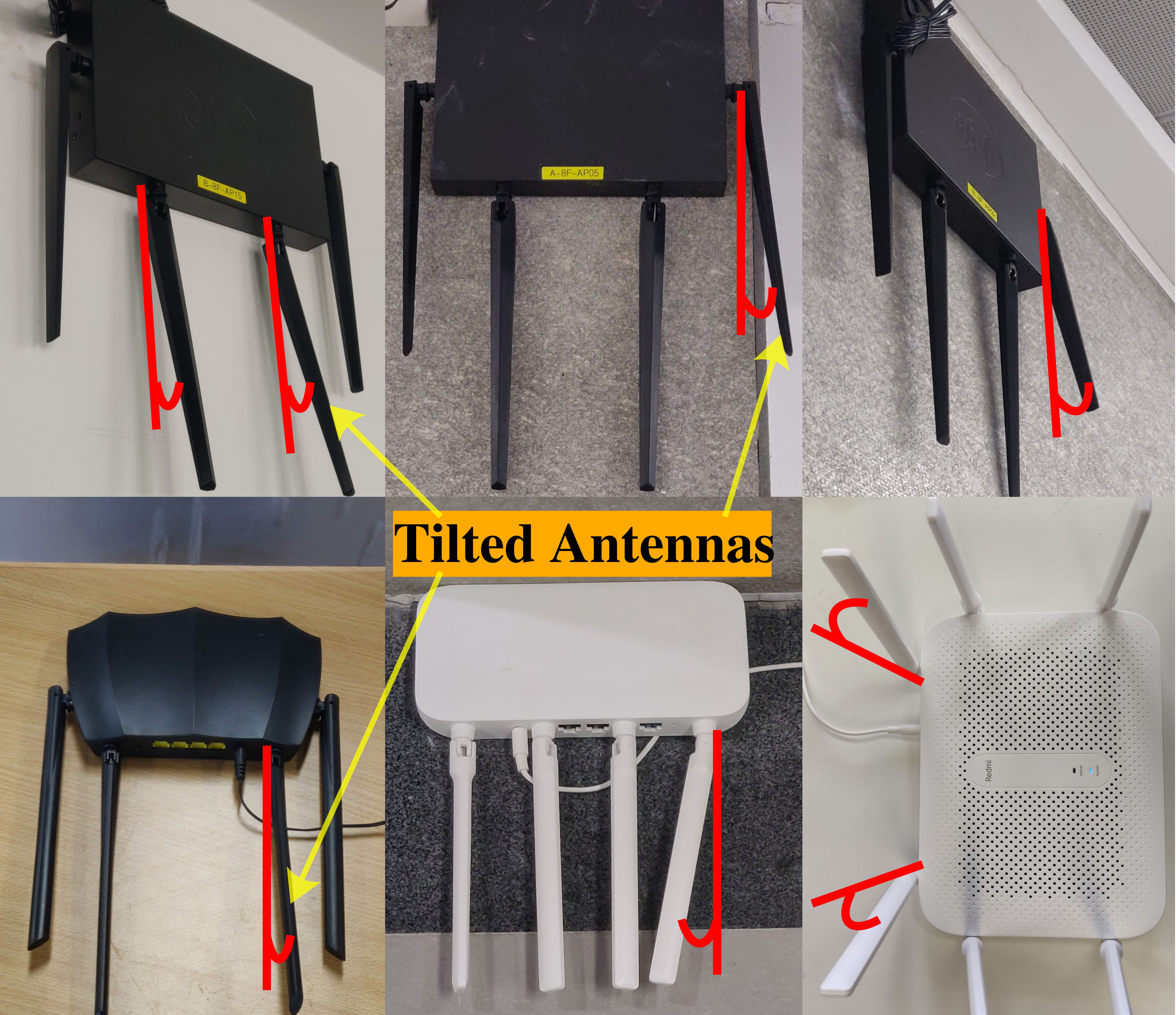}}
	\quad
	\quad
	\subfloat[\label{fig:orientation error}]{
		\includegraphics[width=0.4\linewidth]{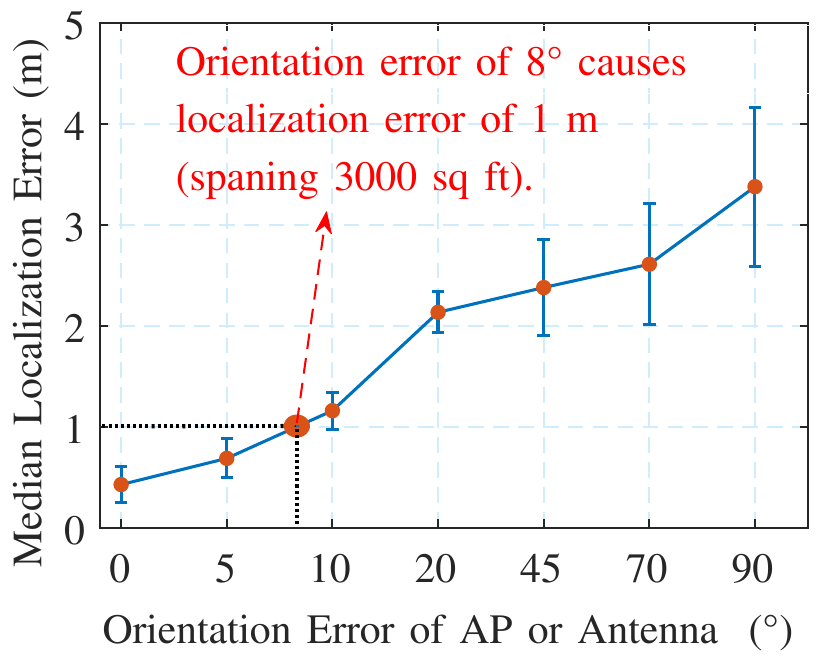}}
    \vspace{-0.18cm} 
	\caption{\textbf{Motivation:} (a) Four orientation errors of AP or antennas (yaw, roll, pitch, irregular tilt). (b) The antennas have different tilted angles in real scenarios. (c) The localization error \textit{vs.} the orientation error of AP or antenna.}\label{fig:introduction_1}
	\vspace{-0.4cm} 
\end{figure}

In particular, we take WiFi localization as the case and analyze the impact of WiFi infrastructure credibility on accuracy. In WiFi localization, antenna orientation is an often overlooked item. However, errors in antenna orientation reduce the accuracy of the localization system~\cite{pelka2018impact,vtc2019antennas,10.5555/3400306.3400325,locap2020}. In real scenarios, as shown in Fig.~\ref{fig:antenna rotation} and Fig.~\ref{fig:AP antenna}, due to manual measurement errors during deployment or the pursuit of higher throughput~\cite{MapFi2021,asus}, the tilt of the AP or antenna orientation is common, and they may have four orientation errors (\ie, yaw, roll, pitch, irregular tilt). It’s worth mentioning that inaccuracies in WiFi AP's or antenna's orientation are more important than inaccuracies in AP location and antenna separation. Specifically, it takes \SI{30}{cm} of AP position drift to introduce \SI{50}{cm} of localization error~\cite{locap2020}, and typically only the exposed antennas’ separations may change. In contrast, according to our test in a scenario that spans 3000 sq ft in area, an orientation error of \SI{8}{^{\circ}} causes a localization error of \SI{1}{m}, as shown in Fig.~\ref{fig:orientation error}.

Therefore, if the orientations of all the the APs' antennas can be accurately measured, this is expected to help the widespread deployment of many WiFi-based localization systems in the real-world and maintain high accuracy for a long time. For example, we regularly report the antennas' orientations of all WiFi APs in the scene, which can guide calibration strategies of antennas. Specifically, for the three situations in which the APs are tilted as a whole (yaw, roll, pitch), the localization systems use the actual reported antenna orientations to locate the user, while the situation where the antennas are irregularly tilted requires less manpower compared to aimless calibration.
To achieve this goal, such a system for measuring the antennas orientations of WiFi APs should satisfy the following three requirements: 
\begin{itemize}
	\item \textbf{Non-invasive measurement:} The system should be able to measure the antennas' orientations of WiFi APs in an unknown physical map. Specifically, there is no need for prior data of antenna, no need for the AP to carry sensors such as gyroscope and camera, and no need for hardware or firmware modifications to AP.
	\item \textbf{Accurate antennas' orientation:} We aim to achieve the localization error of less than \SI{1}{m} in a scenario that spans 3000 sq ft in area. Thus, based on the effects of the antenna orientation errors on the localization accuracy shown in Fig.~\ref{fig:orientation error}, the antenna orientation prediction error of the system should be $\leq \SI{ 8}{^\circ}$. 
	\item \textbf{Low time-cost:} The system should be able to measure the orientations of APs' antennas in as short a time as possible, especially if there are thousands of APs in the physical map, at least not exceeding the manual measurement time. For example, according to our survey, the time for manually measuring four antennas' orientations of one AP is at the minute-level. 
\end{itemize} 

However, to the best of our knowledge, no system exists that satisfies all of the above requirements. There are several methods to measure antenna orientation or tilt angle: \emph{(\romannumeral1)} Manual-based, which is time-consuming, labor-intensive, and subject to human error~\cite{compass}; \emph{(\romannumeral2)} Sensor-based, which requires them to be installed on the antennas~\cite{cpi-2.4aebp,ngabo20183d}; \emph{(\romannumeral3)} Vision-based, which requires sufficient lighting for the antennas to be observed~\cite{yang2022novel}; \emph{(\romannumeral4)} WiFi-based, where some recent works provide high-precision acquisition of APs orientations based on WiFi signal, but they can only measure orientations of APs in the horizontal plane (Fig.~\ref{fig:antenna rotation} (\textit{i}))~\cite{locap2020,MapFi2021}.

In this paper, we propose \oursystem \  that measures the orientations of each of WiFi APs' antennas in the physical space non-invasively, accurately and efficiently. \oursystem's feasibility for antennas orientations measurement stems from the fact that antennas have different radiation or reception capabilities for different directions in the physical space~\cite{friis1946note,262044}.
However, there are three main challenges:

\emph{(\romannumeral1)} All antenna parameters of the electromagnetic wave propagation cannot be obtained. Since we do not have the prior data of the AP antenna and cannot modify the AP, we cannot obtain antenna parameters such as antenna gain and polarization mismatch factor, that is, we cannot directly measure the antenna orientation based on the existing model.

\emph{(\romannumeral2)} In wireless links, the received power is also affected by propagation distance, environment and etc. Since we do not accurately know the environment in advance, the propagation distance of the signal in the medium and the attenuation of other items are unknown, so it is difficult to directly calculate the law that determines how the received power of the antenna varies with all orientations.

\emph{(\romannumeral3)} To the best of our knowledge, we are the first WiFi-based system to measure the orientations of each of AP's antennas in physical space, and doing this at a time cost less than the manual time cost (\eg, minute-level) is challenging.

\begin{figure*}[t]
 	\centering
	\subfloat[\label{fig:friis}]{
		\includegraphics[width=0.19\linewidth]{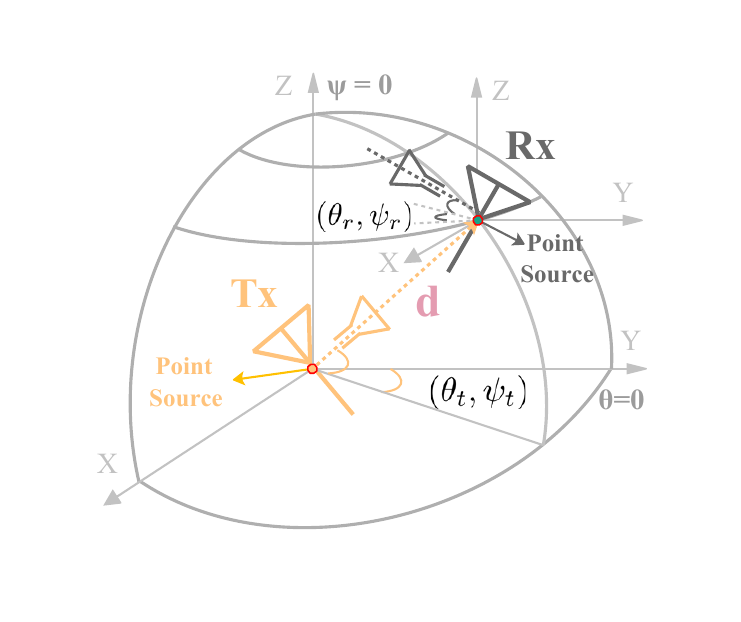}}
 \quad
	\subfloat[\label{fig:radiation}]{
		\includegraphics[width=0.175\linewidth]{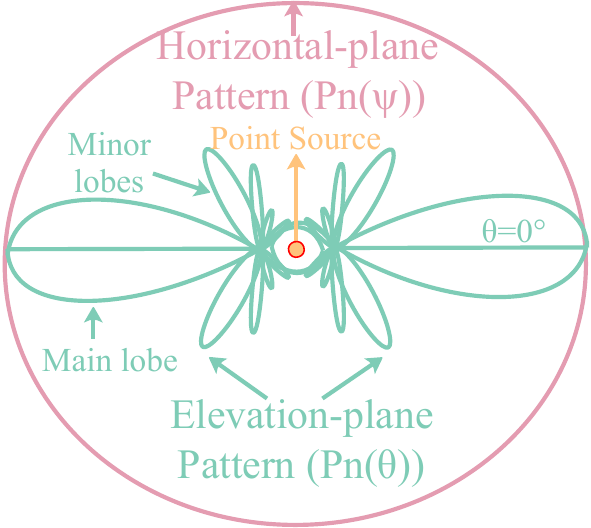}}
 \quad
	\subfloat[\label{fig:pattern}]{
		\includegraphics[width=0.27\linewidth]{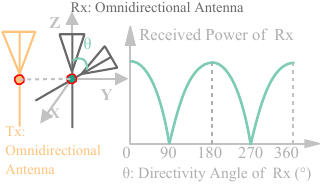}}
 \quad
	\subfloat[\label{fig:polarization}]{
		\includegraphics[width=0.21\linewidth]{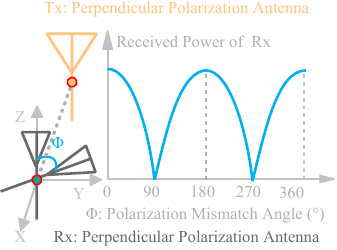}}
        \vspace{-0.25cm}
	\caption{\textbf{Antenna radiation or reception:} (a) Antenna system for Tx-Rx pair. (b) Radiation pattern of omnidirectional antenna. (c) Effect of directivity on received power. (d) Effect of polarization on received power.}
	\label{fig:antenna}
	\vspace{-0.25cm} 
\end{figure*}

\begin{figure*}[t]
	\centering
	\includegraphics[width=0.94\linewidth]{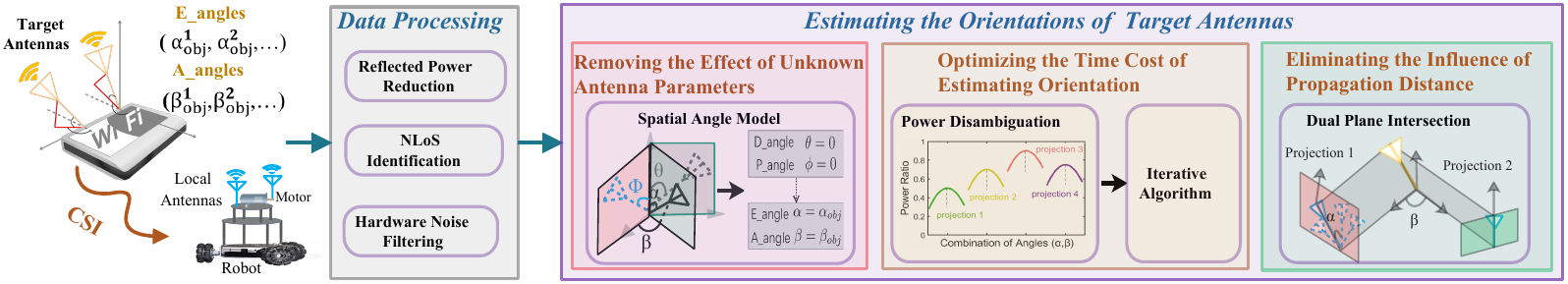}
	\vspace{-0.36cm}
	\caption{\textbf{System overview:} platform, data processing, estimating the orientations of target antennas.}
	\label{fig:system}
	\vspace{-0.3cm} 
\end{figure*}

\textbf{Our solutions.} 
The key idea comes from the fact that there is a mapping between the electric field angles of the Tx-Rx antenna pair and the spatial angles of the local antennas (\textit{i.e.}, the antennas used in \oursystem). \emph{(\romannumeral1)} Firstly, considering that received power is maximized when the local antenna and AP antenna are perfectly aligned, we construct a spatial angle model, which calculates the law of the received power with the local antenna spatial angle. This eliminates the effects of unknown antenna parameters. \emph{(\romannumeral2)} Secondly, we disambiguate the received power according to the orthogonality of antenna directivity and polarization, and construct vertical planes perpendicular to ~\emph{line of sight} (LoS) paths. Specifically, we first obtain several local maxima only on these vertical planes, and then combine to obtain the global maximum to obtain the AP antenna orientation, thereby optimizing the time cost. We adopt an iterative algorithm to further improve the efficiency. \emph{(\romannumeral3)} Thirdly, we describe the AP antenna as a 3-D vector, and construct horizontal planes based on the AP antenna and its projection in the vertical plane. We use the intersection of two horizontal planes to determine the AP antenna orientation, which eliminates the influence of propagation distance between vertical planes. We further improve the accuracy by removing the effect of~\emph{non line of sight} (NLoS) and filtering out ambient noise.

\textbf{Contributions:}
\begin{itemize}
	\item We propose \oursystem, to the best of our knowledge, the first attempt to measure the orientation of each antenna of AP in physical space based on WiFi signals. The advantage of \oursystem \ is that the orientation of each antenna can be measured without prior data and without hardware/firmware modifications to APs.
	\item We design an optimization algorithm combining received power disambiguation with an iterative estimation process, which reduces the sampling points by hundreds of times, thus greatly improving the efficiency. We also build a dual plane intersection model to remove the influence of propagation distance, which improves the accuracy.
	\item We implement \oursystem \ based on a WiFi ~\emph{network interface card} (NIC) combined with a ~\emph{simultaneous localization and mapping} (SLAM) robot. We test our proposed model and techniques in the real world, for different antenna types, geometries and environments, to obtain median errors of both elevation and azimuth angles below \SI{8}{^\circ}. Furthermore, we demonstrate the effectiveness of \oursystem \ through case studies comparing state-of-the-art reverse localization and user localization systems.
\end{itemize}

%% file: section/preliminary.tex
\section{Preliminaries} \label{sec:preliminary}
\subsection{Friis Transmission Formula}
The \textit{Friis transmission formula}~\cite{friis1946note} is used to determine the power received by a lossless and load-matched antenna in a radio communication link~\cite{antenna_third,bevelacqua2021friis}:
\begin{equation}
	P_r = \frac{P_t G_t G_r \lambda^2}{(4 \pi d)^2}, 
	\label{con:Friis}
\end{equation}
where $P_t$ and $P_r$ are the power of Tx and Rx antennas, $G_t$ and $G_r$ are the gains of Tx and Rx antennas, $\lambda$ is the wavelength, and $d$ is the distance between two antennas. It is worth noting that this formula applies to the following conditions~\cite{antenna_third,bevelacqua2021friis}:
\textit{(i)} $d \gg \lambda$, \textit{i.e.} one antenna must be in the far field of the other. \textit{(ii)} The antennas are correctly aligned and have the same polarization. \textit{(iii)} The antennas are in free space, with no multipath. \textit{(iv)} Directivities are both for isotropic radiators.

\subsection{Antenna Directivity}
Fig.~\ref{fig:friis} depicts a point radiation source expressed in spherical coordinates in free space, where the center of the sphere is the antenna phase center~\cite{antenna_third}. However, antenna usually has directivity, and its radiation space is not uniform. A power pattern is a 3-D quantity that describes power as a function of the spherical coordinates $\theta$ and $\psi$~\cite{jordan1968electromagnetic,cheng1989field}. Fig.~\ref{fig:radiation} depicts the radiation pattern of an omnidirectional antenna. The horizontal-plane pattern shows the uniform radiation of \SI{360}{^\circ}. The elevation-plane pattern shows a beam with a certain width, which has the maximum radiation along the $\theta = \SI{0}{^\circ}$ direction.  Fig.~\ref{fig:pattern} shows the effect of antenna directivity on received power, and the received power is maximized when the main lobes of the two antennas are in the same direction.


\subsection{Antenna Polarization Matching}
Polarization is an important indicator of antenna, which describes the trajectories of electric and magnetic field vectors as electromagnetic waves propagate in space \cite{7055373,262044}. When Tx and Rx antennas have the same polarization direction, the received signal is the strongest, which is polarization matching.
Fig.~\ref{fig:polarization} shows the effect of antenna polarization on received power, and the received power is maximized when the two antennas are parallel. For linear polarization, the polarization mismatch factor of the power is~\cite{antenna_third}:
\vspace{-0.1cm} 
\begin{equation}
F = cos^2 \phi,
\label{con:polar}
\vspace{-0.15cm} 
\end{equation}
where $\phi$ is the difference of inclination between Tx and Rx.

%% file: section/overview.tex
\section{Overview} \label{sec:overview}

\subsection{Problem Statement}\label{sec:problem}
As shown in the left part of Fig.~\ref{fig:system}, we refer to the antennas of WiFi AP to be estimated as target antennas, and the antennas used in \oursystem \ as local antennas. We first briefly introduce the problem of estimating target antenna's orientation in physical space, including the elevation angle (E\_angle) $\alpha^i_{obj}$ and azimuth angle (A\_angle) $\beta^i_{obj}$, where $i$ is the reference number of the target antennas. In our research, we assume that the location of each target antenna is known, which can be determined using a method like LocAP~\cite{locap2020}. Note that the errors of antenna separation in LocAP increase greatly when the target antennas are tilted. But the errors have no effect on our research, because we are concerned with the true location of the target antenna (especially the antenna element). Specifically, based on the determined location of each target antenna, \oursystem \ takes state-series WiFi signals \pmb{$\mathcal{H}$} = $\{\textbf{H}_{\alpha_i,\beta_k} \ | \  \alpha_i \in [0,2 \pi) , \beta_k \in [0,2 \pi) \}$ from the local antennas during several different states (combinations of E\_angle and A\_angle $(\alpha_i,\beta_k)$) as input, and then derives the orientations of all target antennas $(\alpha^1_{obj}, \beta^1_{obj}, \alpha^2_{obj}, \beta^2_{obj}, \cdots)$.

\begin{figure*}[t]
	\centering
	\subfloat[\label{fig:friis_model}]{
		\includegraphics[width=0.25\linewidth]{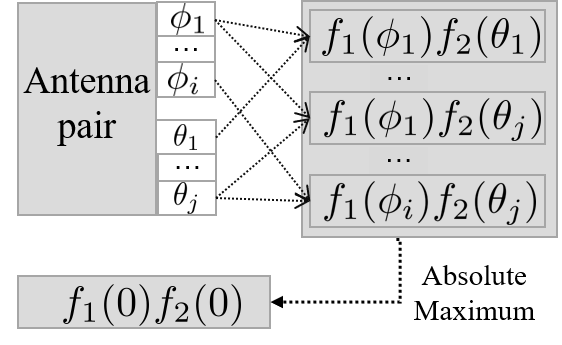}}
	\quad
	\quad
	\subfloat[\label{fig:convert}]{
		\includegraphics[width=0.27\linewidth]{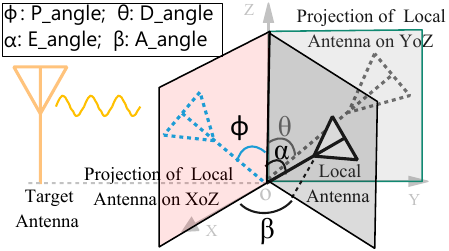}}
	\quad
	\quad
	\subfloat[\label{fig:basic_model}]{
		\includegraphics[width=0.24\linewidth]{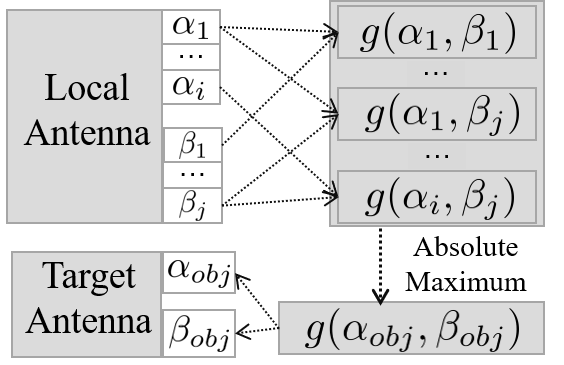}}
	\vspace{-0.2cm}
	\caption{\textbf{Unknown antenna parameters effect removing:} (a) For a target-local antenna pair, the received power can be expressed as the function of $\phi$ and $\theta$, and reaches the maximum when ($\phi=0,\theta=0$). (b) Through coordinate transformation, we convert the unknown ($\phi,\theta$) of the antenna pair to the known ($\alpha,\beta$) of the local antenna; note that ($\alpha=\alpha_{obj}, \beta=\beta_{obj}$) corresponds to ($\phi=0,\theta=0$). (c) For a target antenna with fixed orientation, the received power can be expressed as the function of $\alpha$ and $\beta$ of the local antenna, and reaches the maximum when ($\alpha=\alpha_{obj},\beta=\beta_{obj}$).}
	\label{fig:model}
	\vspace{-0.3cm}
\end{figure*}

\subsection{\oursystem \ Architecture}\label{sec:system_overview}
In this paper, the target antennas are used as the Tx antennas, and the local antennas are used as the Rx antennas.

\textbf{Data processing.} We control \oursystem \ to collect~\emph{channel state information} (CSI) between target antennas and local antennas. Then, we remove CSI in NLoS, and use dual antenna to filter hardware noise to extract stable received power.

\textbf{Removing the effect of unknown antenna parameters ($\phi, \theta$).} Specifically, we reconstruct the \textit{Friis transmission formula} to a spatial model, that measures the received power according to the known spatial angles ($\alpha, \beta$) of the local antennas. The target antenna's orientation is obtained by solving for the maximum of the received power.

\textbf{Optimizing the time cost of estimating orientation.} We construct the vertical and horizontal planes according to the LoS path of the antenna pair, obtain the antenna orientation by first getting several local maxima only on these vertical planes and then getting their global maximum, and use an iterative algorithm to further improve the efficiency.

\textbf{Eliminating the influence of propagation distance.} We use the spatial geometry principle that the intersection of two planes is unique line to remove the influence of propagation distance while making the system easier to deploy.

%% file: section/design.tex
\section{System Design} \label{sec:system}
\subsection{Removing the Effect of Unknown Antenna Parameters}
\textbf{Estimating antenna's orientation based on antenna parameters.}
To estimate the orientation of the target antenna, we construct an antenna system consisting of the target antenna and the local antenna. According to Equ.~\ref{con:Friis} and Equ.~\ref{con:polar},  when considering antenna directivity and polarization, the received power $P_r$ of the local antenna:
\vspace{-0.25cm} 
\begin{equation}
	P_r = k cos^2 \phi \cdot \frac{P_t G(\theta_t) G(\theta_r) \lambda^2}{(4 \pi d)^2 },
	\label{con:Friis2}
	\vspace{-0.15cm} 
\end{equation}
where $k$ is the antenna efficiency factor, which is a constant, $\phi$ is the polarization mismatch angle, $P_t$ is the transmit power of the target antenna, $G(\theta_t)$ and $G(\theta_r)$ are the gains of target antenna and local antenna, $\lambda$ is the wavelength, and $d$ is the distance between two antennas. It is obvious that the received power changes due to the change of D\_angle ($\theta_t,\theta_r$) and P\_angle $\phi$. As a result, when we obtain the received power $P(\phi, \theta)$, the relative angle between the target antenna and the local antenna is:
\vspace{-0.2cm} 
\begin{equation}
	f(\phi,\theta) = \frac{P(\phi,\theta) \ d^2}{P_{t} \ \lambda^2 } = f_1(\phi) f_2(\theta),
    \label{con:ideal_model}
    \vspace{-0.15cm} 
\end{equation}
where $\theta$ is the relative directivity angle between the target antenna and local antenna, which is independent of $\phi$, $f(\phi,\theta)$ is the product of two terms respectively depending on $\phi,\theta$ and reaches the absolute maximum when the two antennas are perfectly aligned (\textit{i.e.}, $\phi = 0, \theta = 0$); see also Fig.~\ref{fig:friis_model}.

\textbf{Estimating antenna's orientation based on spatial angles.}
Noting that $\phi$ and $\theta$ are unknown, the mapping $f(\theta,\phi)$ is not available. When the spatial angle ($\alpha_{obj}, \beta_{obj}$) of the target antenna is constant and the spatial angle ($\alpha, \beta$) of the local antenna is known, we can convert the unknown angles ($\phi, \theta$) to the known angles ($\alpha, \beta$), as shown in Fig.~\ref{fig:convert}. We then establish the mapping of the spatial angle of the local antenna to the received power:
\vspace{-0.2cm} 
\begin{equation}
	g(\alpha,\beta) = \frac{P(\alpha,\beta) \ d^2}{P_{t} \ \lambda^2 }, \alpha \in [-\frac{\pi}{2}, \frac{\pi}{2}), \beta \in [-\frac{\pi}{2}, \frac{\pi}{2}),
    \label{con:signal_model}
    \vspace{-0.15cm} 
\end{equation}
where $P(\alpha,\beta)$ is the received power when E\_angle is $\alpha$ and A\_angle is $\beta$ of the local antenna, $g(\alpha,\beta)$ is the function related to $\alpha$ and $\beta$, and reaches the absolute maximum when the two antennas are perfectly aligned (\textit{i.e.}, $\alpha=\alpha_{obj}, \beta=\beta_{obj}$), as shown in Fig.~\ref{fig:basic_model}.

\textbf{Complexity analysis.}
In theory, we can traverse the orientation of the local antenna in physical space to obtain the received power and get a set of $g$  values. However, this method needs to collect a large amount of data. For example, the time cost to obtain one $g$ is \SI{1}{s}, which includes the acceleration time of the motor, the time of collecting and processing CSI. Here, each step angle is set to \SI{2}{^\circ}. The time cost to traverse is $(180/2)^3 = \SI{72900}{s} = \SI{202.5}{h}$, which is unacceptable. Thus, we need to optimize the time cost so that the system can quickly estimate the target antenna orientation.

\begin{figure*}[t]
	\centering
	\subfloat[]{
	    \label{fig:ambiguation}
	    \includegraphics[width=0.27\textwidth]{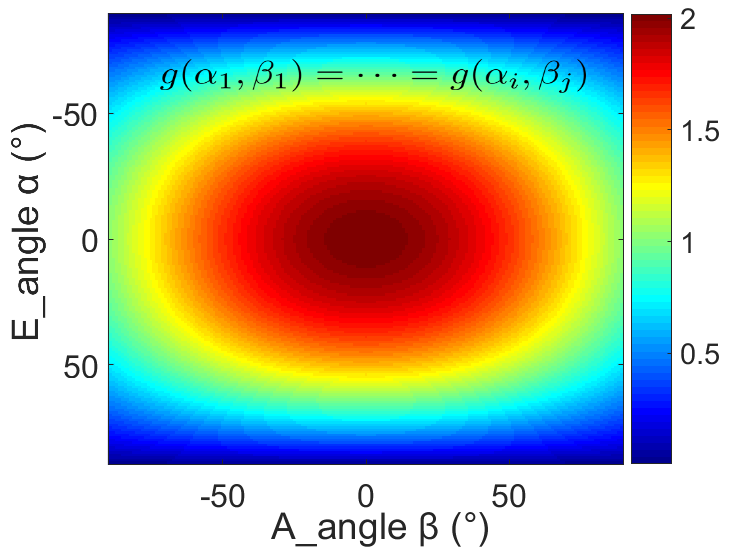}}
	\quad
	\subfloat[]{
	    \label{fig:model_3D}
	    \includegraphics[width=0.35\textwidth]{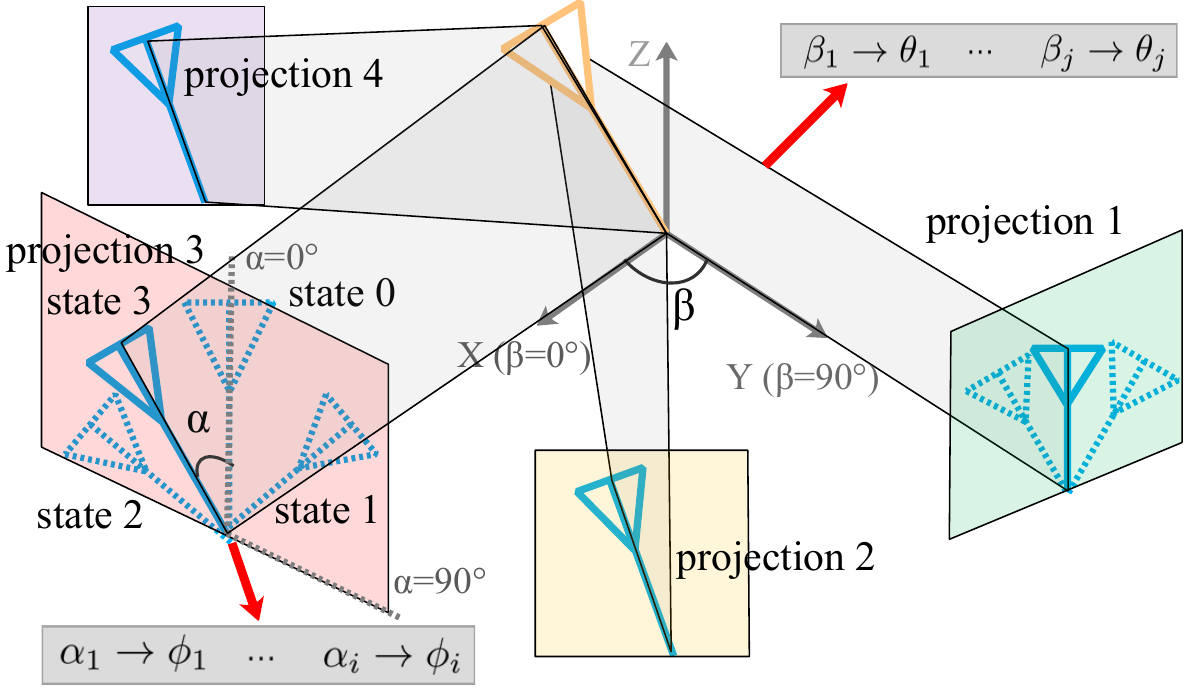}}
	\quad
	\subfloat[]{
	    \label{fig:power_max}
	    \includegraphics[width=0.22\textwidth]{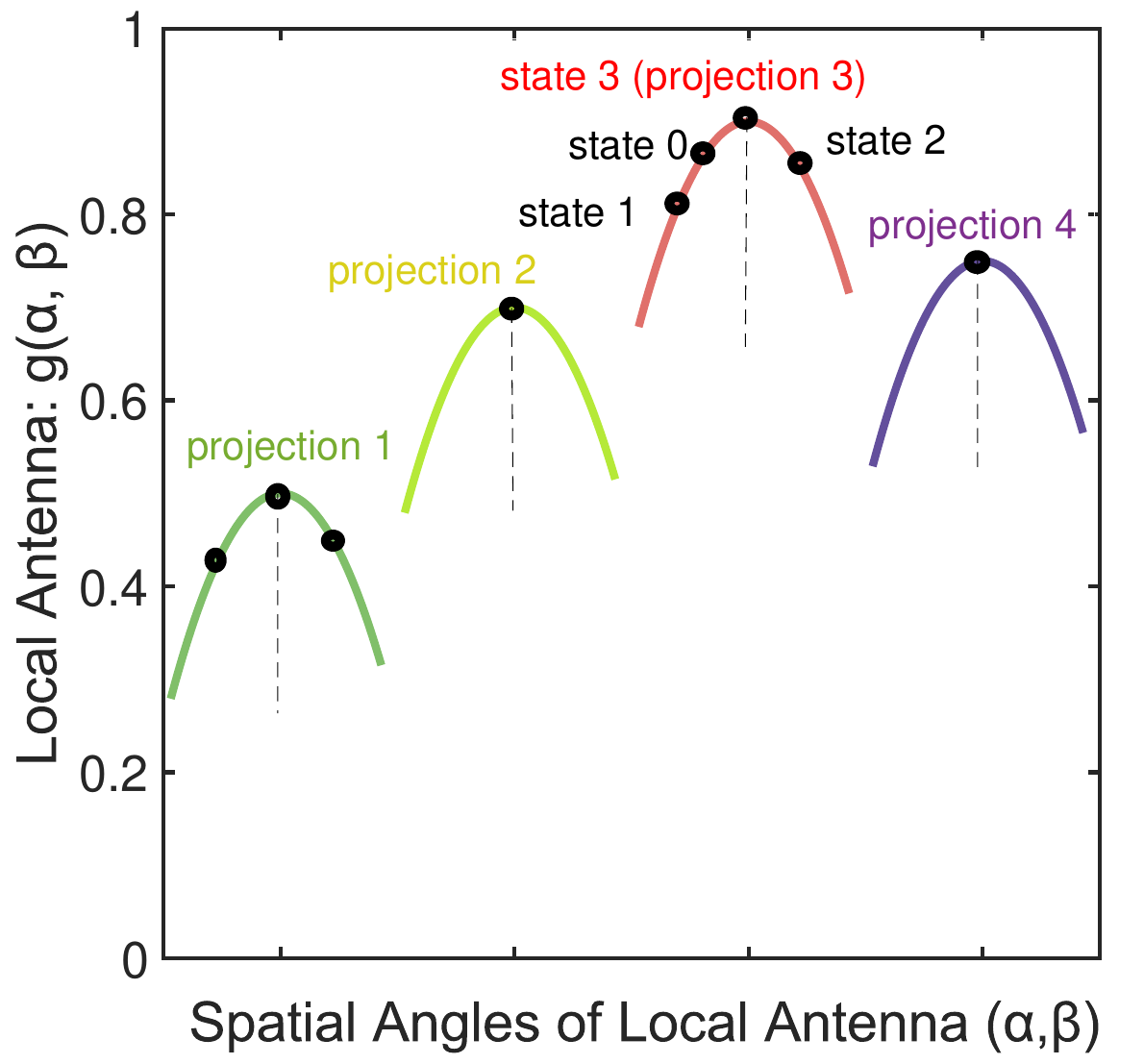}}
	\vspace{-0.18cm}
	\caption{\textbf{Time cost optimization and propagation distance influence removing:} (a) The local antenna exhibits the same received power at different ($\alpha,\beta$). (b) Using the orthogonality of $\phi$ and $\theta$, we convert $\alpha$ and $\beta$ to $\phi$ and $\theta$ respectively, and remove the influence of propagation distance based on the geometric principle of plane intersection. (c) The orientation of the target antenna can be quickly obtained by solving for the relative maximum of each vertical plane and then for their absolute maximum.}\label{fig:search model}
	\vspace{-0.3cm}
\end{figure*}

\subsection{Optimizing the Time Cost of Estimating Orientation}
\textbf{Received power disambiguation.}
To reduce the time cost, we first disambiguate the received power. As shown in Fig.~\ref{fig:ambiguation}, multiple different $(\alpha,\beta)$ exhibit the same $(\phi,\theta)$, resulting in the same received power, which leads to ambiguity. The reason is that $\alpha$ and $\beta$ act together on the received power, \textit{i.e.}, they are correlated. 
Noting that $\phi$ and $\theta$ are independent of each other, we can convert angles of ($\alpha  \rightarrow  \phi, \quad \beta  \rightarrow  \theta$) for received power disambiguation.

We then describe the process of antenna orientation estimation after angle conversion. As shown in Fig.~\ref{fig:model_3D}, we construct multiple mutually perpendicular planes based on the LoS path from the target antenna to the local antenna. We call the vertical plane ($\Pi_{\beta_1},\Pi_{\beta_2},\cdots$) perpendicular to the LoS path and the horizontal plane ($\Pi_{1},\Pi_{2},\cdots$) parallel to the LoS path. First, we rotate the local antenna in one certain vertical plane $\Pi_{\beta_k}$, and the E\_angle is $\alpha$. This process is carried with fixed ($\theta = \theta_{k}$), thus the 
function $g$ is only related to $\phi$ ($\alpha \rightarrow  \phi$). Hence, in plane $\Pi_{\beta_k}$, we obtain:
\vspace{-0.1cm} 
\begin{equation}
	g_{1_k}(\alpha) = g(\alpha, \beta) |_{\beta=\beta_k} = \frac{P(\alpha) d^2}{f_2 |_{\theta=\theta_k} P_t \lambda^2}, \  k \in [1,K],
    \label{con:search_alpha}
    \vspace{-0.1cm} 
\end{equation}
where $K$ is the number of vertical planes. As shown in Fig.~\ref{fig:power_max}, we get a E\_angle $\alpha_{\beta_k}$ corresponding to the relative maximum of $g_{1_k}(\alpha)$ in plane $\Pi_{\beta_k}$:
\vspace{-0.15cm} 
\begin{equation}
	\alpha_{\beta_k} = \mathop{\arg \max}_{\alpha} \ g_{1_k}(\alpha),  
    \label{con:max_alpha}
    \vspace{-0.1cm} 
\end{equation}
where $g_{1_k}(\alpha)$ reaches the relative maximum only when $\phi = 0$, and $\alpha = \alpha_{\beta_k}$ at this time. Second, for the relative maximum of each vertical plane, we fix $\phi$ ($\phi = 0$), so the function $g$ is only related to $\theta$ ($\beta \rightarrow  \theta$). Hence, in plane $\Pi_{k}$, we can obtain the following:
\vspace{-0.15cm} 
\begin{equation}
	g_{2_k}(\beta) = g(\alpha, \beta) |_{\alpha=\alpha_{\beta_k}} = \frac{P(\beta) d^2}{f_1 |_{\phi=0} P_t \lambda^2}, \  k \in [1,K].
    \label{con:search_beta}
    \vspace{-0.1cm} 
\end{equation}
As shown in Fig.~\ref{fig:power_max}, we can measure an absolute maximum of $g_{2_k}(\beta)$ in planes $\Pi_{k}$ and get the angle:
\vspace{-0.1cm} 
\begin{equation}
\begin{aligned}
        \beta_{obj} = \mathop{\arg \max}_{\beta} \ g_{2_k}(\beta), 
    \label{con:max_beta}
\end{aligned}
    \vspace{-0.15cm} 
\end{equation}
where $g_{2_k}(\beta)$ reaches the absolute maximum only when $\theta = 0$, and $\alpha = \alpha_{obj},\beta = \beta_{obj}$ at this time. Obviously, the $g(\alpha_{obj}, \beta_{obj})$ is the absolute maximum in different horizontal planes, and is also the absolute maximum in different vertical planes. Therefore, the angle combination $(\alpha_{obj}, \beta_{obj})$ is the orientation of the target antenna in physical space.

\textbf{Iterative algorithm.}
Naively, we can measure $g$ over the range $[-\pi/2, \pi/2)$ of $(\alpha, \beta)$, but the time cost is still huge. For example, traversing the range of $\alpha$ in each vertical plane, and assuming that the time cost to obtain one $g$ is \SI{1}{s}, the steps of $\alpha$ and $\beta$ are \SI{2}{^\circ}, then the time cost is $(180/2+1)^2 = \SI{8281}{s} \approx \SI{138}{min}$.

We further optimize the time cost by reducing the number of vertical planes and the number of measurements in each vertical plane. The observation is that the closer to the perfectly aligned vertical plane, the bigger the projection's E\_angle, as shown in Fig.~\ref{fig:model_3D}. Thus, we can quickly estimate $\alpha_{obj}$ and $\beta_{obj}$ as a whole in an iterative manner. Specifically, we first apply the procedure in Equ.~\ref{con:search_alpha} - Equ.~\ref{con:max_beta} in only two adjacent vertical planes to determine the trend of the target antenna orientation (the average time cost is $((180/2)/2) \times 2 = \SI{90}{s}$), and then move and rotate the local antenna towards this trend according to the state of the previous vertical plane, until the absolute maximum  is attained (the average time is $(180/2)/2 = \SI{45}{s}$). The total time cost is $\SI{135}{s}$. However, the iterative algorithm is very dependent on the measurements of the previous state. The error of $g$ can seriously affect the accuracy of the target antenna orientation. Hence, we need to remove some factors that affect $g$ such as propagation distance $d$, ambient noise.

\subsection{Eliminating the Influence of Propagation Distance}
To remove the influence of propagation distance on $g$, our basic idea is to make the measurements in each vertical plane independent. An important observation is that the projection of the target antenna in each vertical plane corresponds to the relative maximum of $g$ in this vertical plane, as shown in Fig.~\ref{fig:model_3D}. Then we can obtain the target antenna orientation by using two independent vertical planes based on the spatial geometry principle of ``\textit{intersection of two intersecting planes}''. The specific solution is given below.

In the following, we describe both the target antenna and the local antenna as 3-D vectors. The unit direction vector of target antenna is defined as:
\vspace{-0.05cm}
\begin{equation}
    \bm{ \hat e_{d} } = (\sin{\hat \alpha_{obj}^{m}} \cos{\hat \beta_{obj}^{m}}, \sin{\hat \alpha_{obj}^{m}} \sin{\hat \beta_{obj}^{m}}, \cos{\hat \alpha_{obj}^{m}}),
\label{con:direction_vector1}
\vspace{-0.05cm}
\end{equation}
where $m \in [1, \cdots, M]$ and $M$ is the number of target antennas.
The unit normal vector of vertical plane $\Pi_{\beta_k}$ at $\beta_k$:
\vspace{-0.05cm}
\begin{equation}
\begin{aligned}
\vspace{-0.05cm} 
    \bm{e_{n_{\beta_k}}} = (\cos{\beta_k},\sin{\beta_k},0), \quad for \quad k = 1, 2, \cdots, K.
\end{aligned}
\label{con:normal_vector1}
\end{equation}
In plane $\Pi_{\beta_k}$, the E\_angle is $\hat \alpha_{\beta_k}^{m}$ when $P$ reaches its relative maximum, and the local antenna is the projection of the target antenna in the vertical plane, so the unit direction vector of projection is:
\vspace{-0.1cm} 
\begin{equation}
\begin{aligned}
    \bm{\hat e_{d_{\beta_k}}} = (-\sin{\hat \alpha_{\beta_k}^{m}} \sin{\beta_k}, \sin{\hat \alpha_{\beta_k}^{m}} \cos{\beta_k}, \cos{\hat \alpha_{\beta_k}^{m}}).
\end{aligned}
\label{con:direction_vector2}
\vspace{-0.05cm} 
\end{equation}
Thus, the unit normal vector of the plane $\Pi_{k}$ constructed by the target antenna and its projection:
\vspace{-0.1cm} 
\begin{equation}
\begin{aligned}
    \bm{\hat e_{n_k}} &= \bm{\hat e_{d_{\beta_k}}} \times \bm{e_{n_{\beta_k}}}   \\
                 &= (\cos{\hat \alpha_{\beta_k}^{m}} \sin{\beta_k}, - \cos{\hat \alpha_{\beta_k}^{m}} \cos{\beta_k}, \sin{\hat \alpha_{{\beta}_k}^{m}}).
\end{aligned}
\label{con:normal_vector2}
\vspace{-0.05cm} 
\end{equation}
In theory, $\bm{\hat e_{d}}$ should always be perpendicular to $\bm{\hat e_{n_k}}$:
\vspace{-0.05cm} 
\begin{equation}
\centering
    \bm{\hat e_{d}} \bigcdot \bm{\hat e_{n_k}} = 0, \quad for \quad k = 1, 2, \cdots, K.
\label{con:inner_product}
\vspace{-0.05cm} 
\end{equation}
Then, we estimate the E\_angle $\hat \alpha_{obj}^{m}$ and the A\_angle $\hat \beta_{obj}^{m}$ based on two different vertical planes. In addition, we use the iterative algorithm to quickly measure the relative maximum $\hat \alpha^m_{\beta_k}$ in each vertical plane, and the average time cost is $(180/2/2) \times 2 = \SI{90}{s}$. Finally, we use multiple vertical planes to solve the least squares problem to improve the accuracy. 

\begin{figure*}[t]
	\centering
        \subfloat[\label{fig:multipath}]{
    	\includegraphics[width=0.21\textwidth]{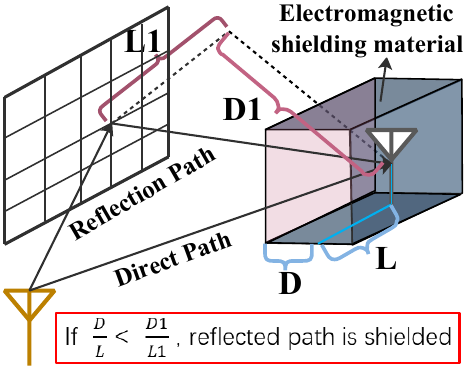}}
	\quad	
        \subfloat[\label{fig:nlos}]{
    	\includegraphics[width=0.318\textwidth]{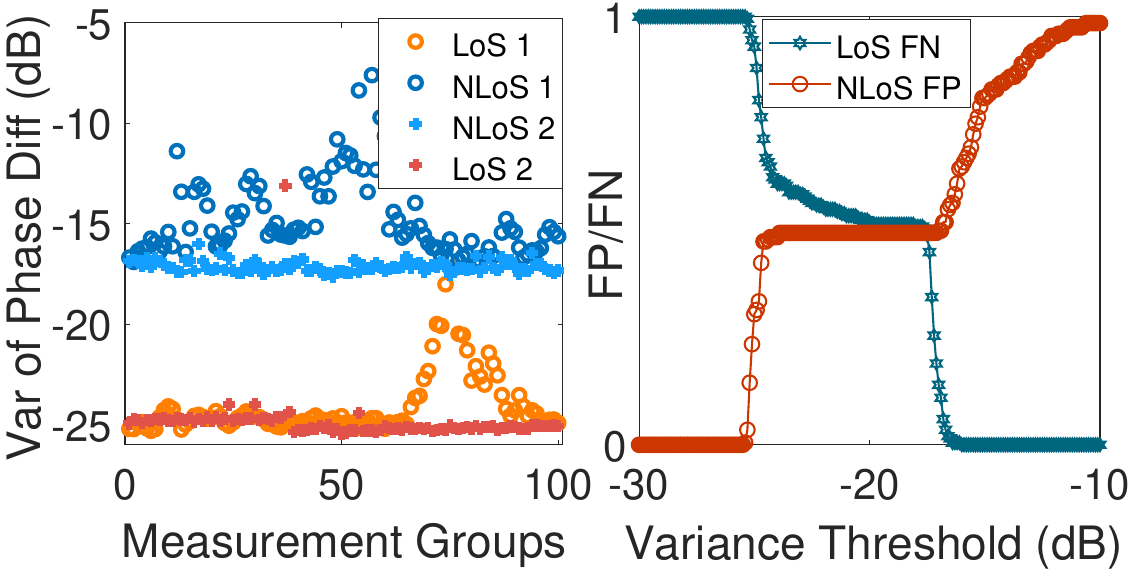}}
	\quad	
	\subfloat[\label{fig:dual_antennas}]{
		\includegraphics[width=0.16\textwidth]{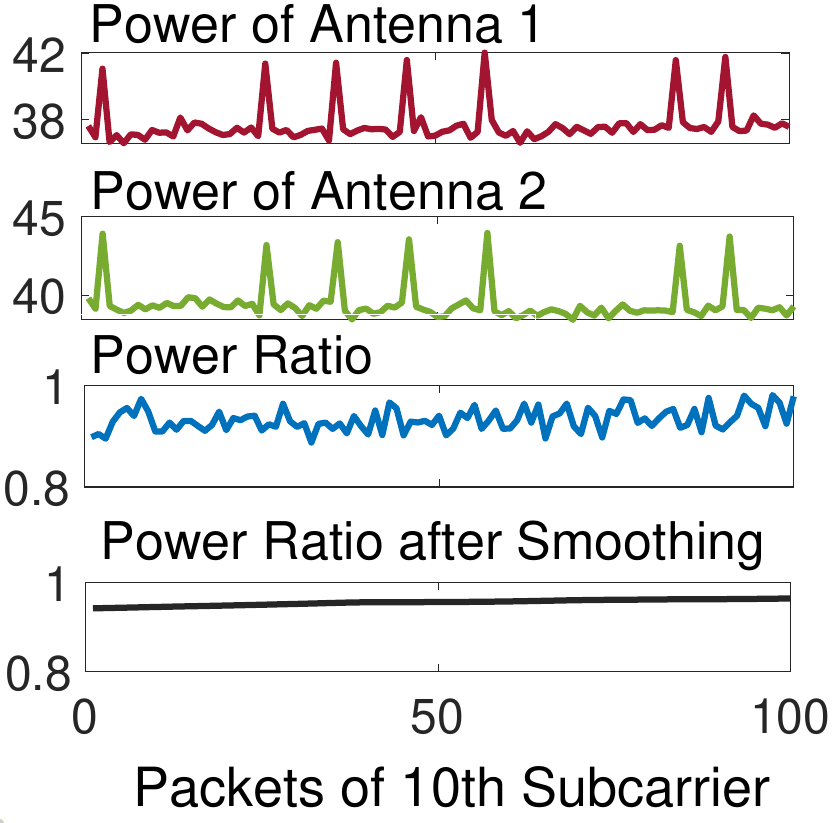}}
	\quad
	\subfloat[\label{fig:subcarrier}]{
		\includegraphics[width=0.165\textwidth]{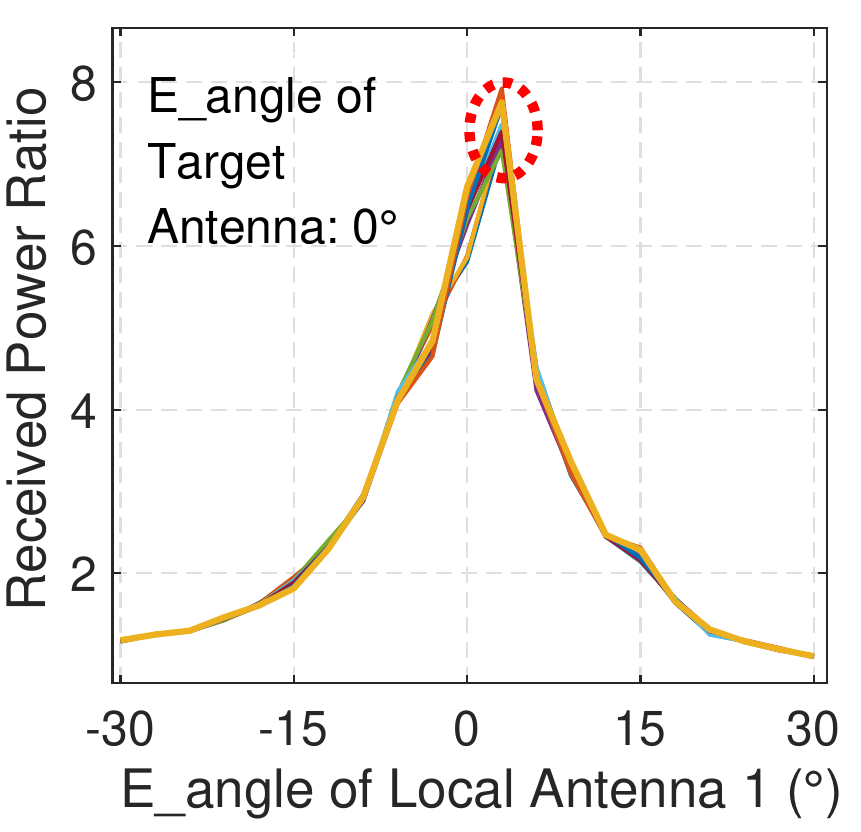}}
	\vspace{-0.2cm}
	\caption{\textbf{Data processing:} (a) We add electromagnetic shielding material around the local antennas to reduce the reflected power since the direction of the direct path is known. (b) LoS/NLoS identification based on dual antenna phase difference variance. (c) Received power ratio after filtering hardware noise by dual antenna ratio and data smoothing. (d) We estimate the E\_angle of target antenna based on the extracted received power ratio.}
	\label{fig:preprocessing}
	\vspace{-0.3cm}
\end{figure*}

\subsection{Data Processing}
\textbf{Reflected power reduction.}
So far we have assumed that there is only LoS path from the WiFi AP to the robot. However, the environment has multipath, and strong multipath can cause significant drop of accuracy for our algorithms due to power measurement distortion. Our observation is that the LoS path is known in our problem as mentioned in Sec.~\ref{sec:problem}, so we can reduce the power of multipath reflections through hardware. Specifically, we added materials around the local antennas, and such materials are readily available and can effectively shield electromagnetic wave signals~\cite{khan2022estimating}. The reflected signal is shielded when the propagation path meets the condition: $\frac{D}{L} < \frac{D_i}{L_i}$, where $D$ and $L$ are the dimensions of the electromagnetic shielding box, $D_i$ and $L_i$ are the distance components of the $i$-th reflection path to the electromagnetic shielding box in two vertical directions, as shown in Fig.~\ref{fig:multipath}.

\textbf{NLoS identification.}
Another factor worth paying attention to is the NLoS, which may lead to large fluctuations in power at the same position~\cite{benedetto2007dynamic,zhou2014lifi}, so that the projection cannot be accurately measured. We utilize a method similar to PhaseU~\cite{wu2015phaseu} to identify LoS/NLoS in real-time. Specifically, NLoS conditions involve more abundant reflections, diffractions, and refractions, so signals traveling through NLoS typically behave more randomly, in terms of amplitude and phase~\cite{wu2015phaseu}. We quantify the difference between LoS and NLoS by incorporating the frequency diversity feature as a weight parameter to compute the spatial phase difference variance of the two antennas. As shown in Fig.~\ref{fig:nlos}, LoS/NLoS can be identified based on certain a threshold, we use two different obstacles to create NLoS conditions (concrete pillar, wooden board). In particular, we set the threshold = $-20dB$, and we treat it as NLoS when the variance is greater than the threshold. Then we optimize the algorithm, if NLoS is identified, the target antenna is not measured at this position, and the next position is used instead.

\textbf{Hardware noise filtering.}
We use an~\emph{industrial personal computer} (IPC) with NIC to collect CSI. However, the accuracy of CSI estimation is affected by hardware noise, such as power control uncertainty error of~\emph{automatic gain control} (AGC) and electromagnetic noise~\cite{niu2021wimonitor}. For~\emph{multi-input multi-output} (MIMO) systems, the noise of multiple antennas at the same sample are approximate, so the dual antenna ratio can be used to filter hardware noise~\cite{10.1145/3351279}. As shown in Fig.~\ref{fig:dual_antennas}, the power ratio is more stable than single antenna. Finally, we extract stable received power based on the above processed CSI, as shown in Fig.~\ref{fig:subcarrier}, which can measure the E\_angle of the target antenna well.

%% file: section/implementation.tex
\begin{figure}[t]
\centering
\begin{minipage}{0.4\linewidth}
    \centering
	\subfloat[\label{fig:hardware}]{
	    \includegraphics[width=0.85\linewidth]{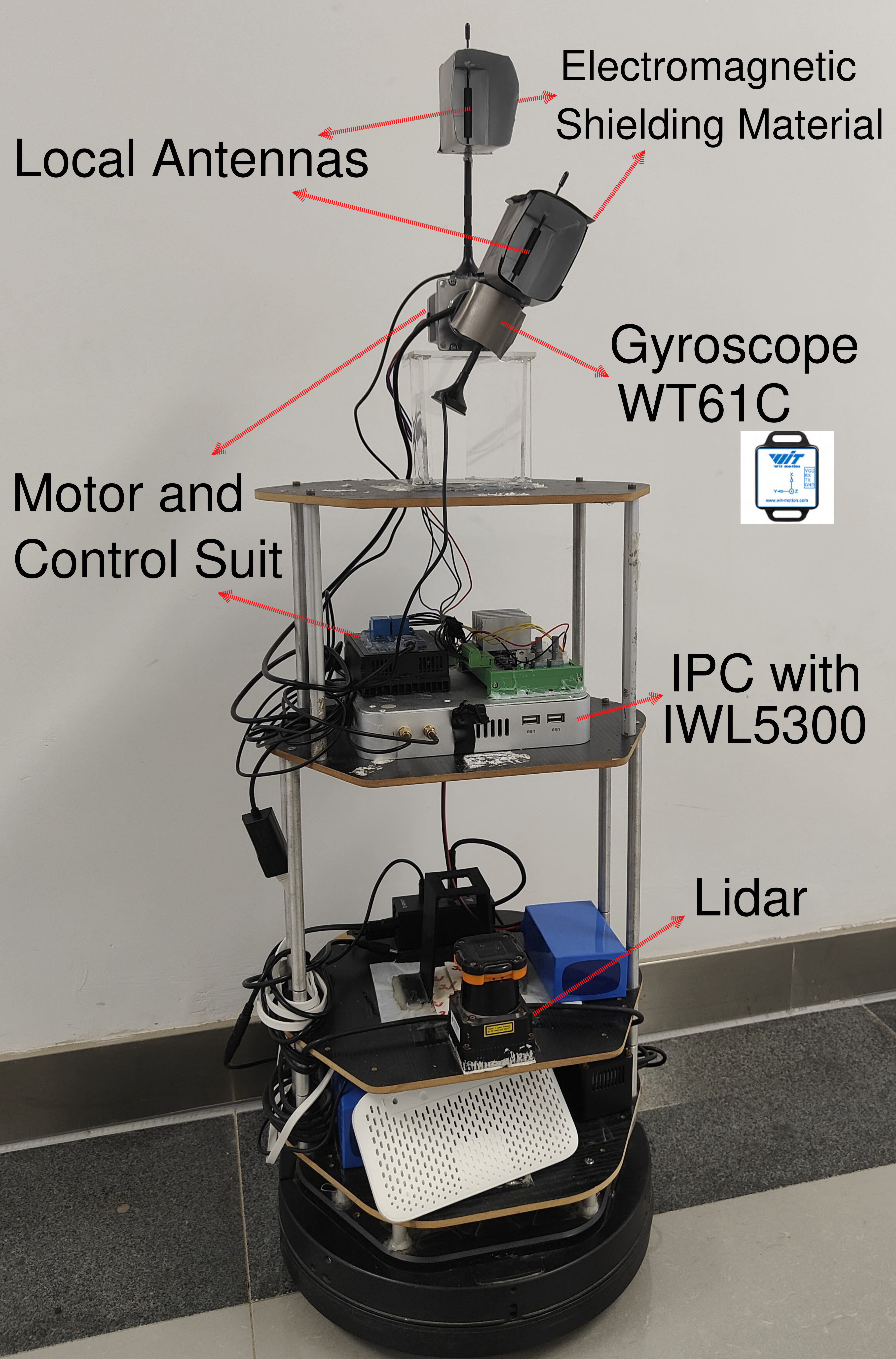}}

    \vspace{-0.4cm}
     \subfloat[\label{fig:bot_error}]{
	    \includegraphics[width=0.85\linewidth]{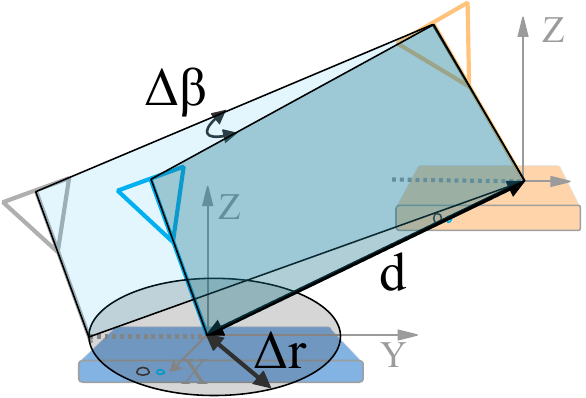}}
\end{minipage}
\vspace{-0.12cm}
\begin{minipage}{0.55\linewidth}
    \subfloat[\label{fig:deployment}]{
    	\includegraphics[width=0.96\linewidth]{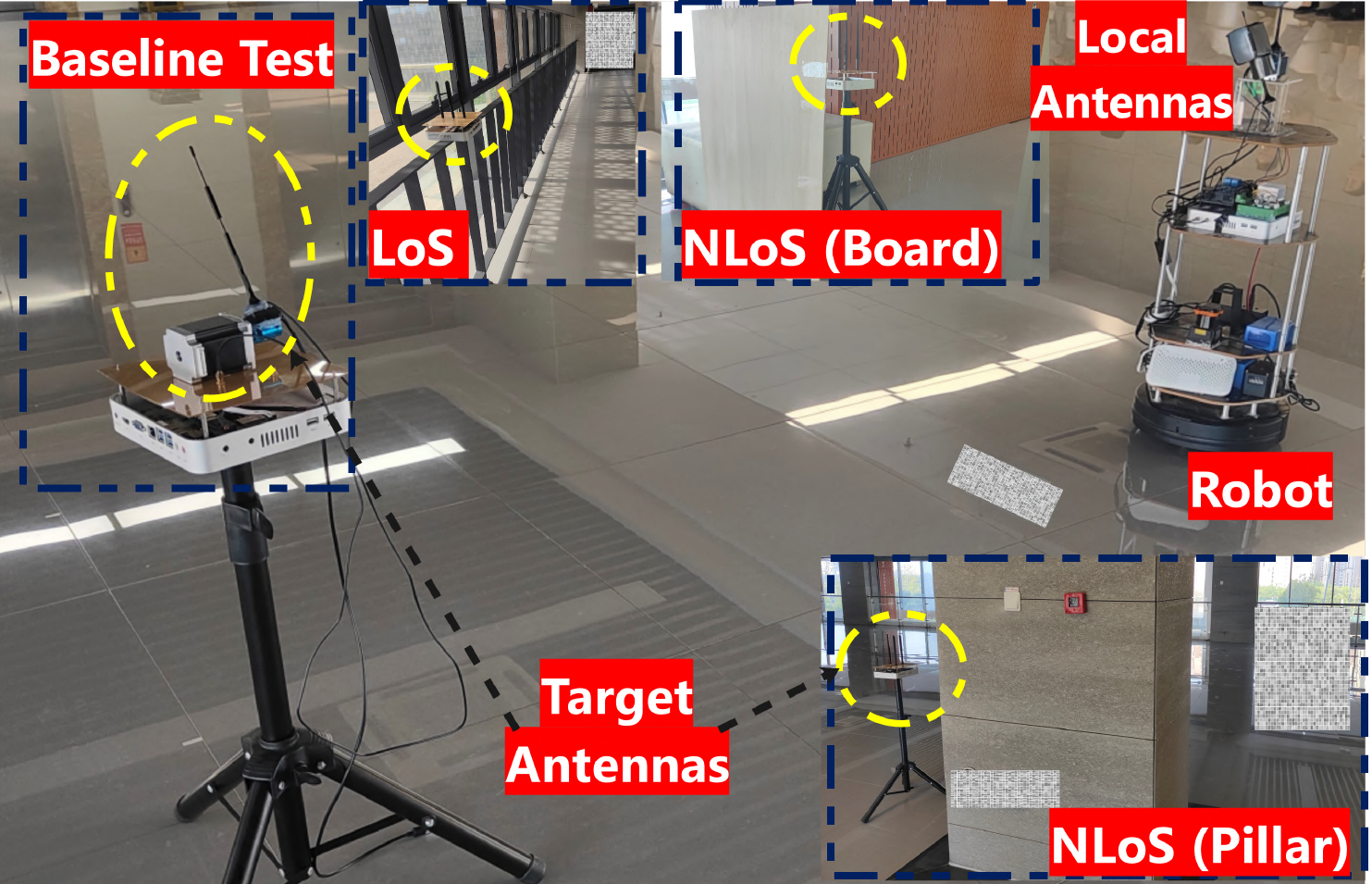}}

    \vspace{-0.3cm}
    \subfloat[\label{fig:hall}]{
	    \includegraphics[width=0.96\textwidth]{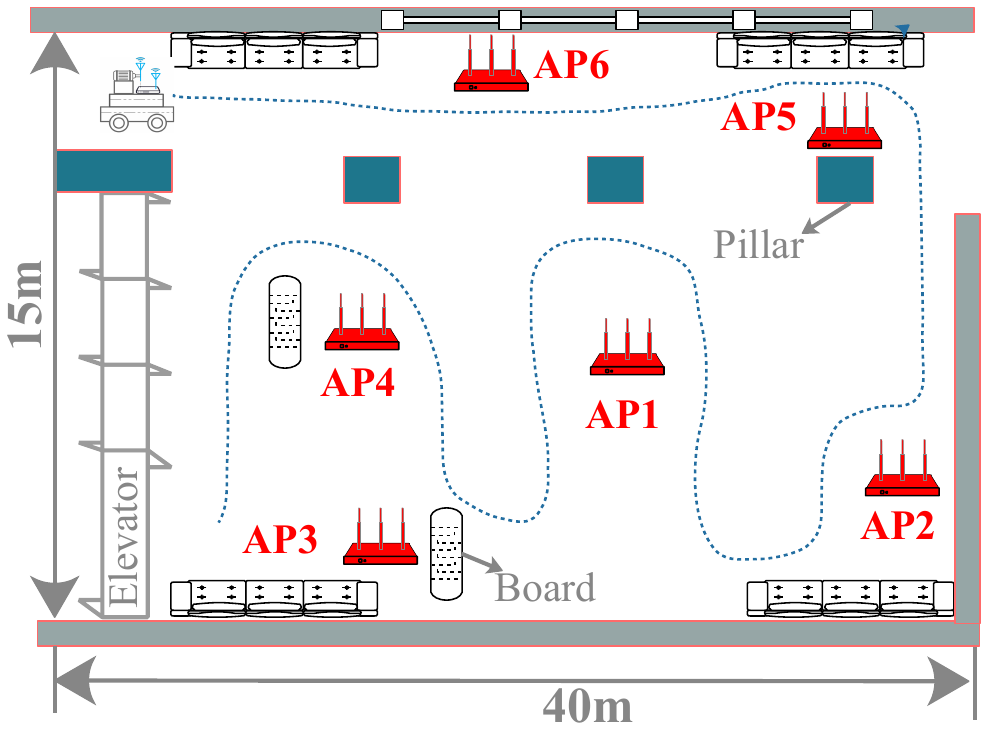}}
\end{minipage}
    \vspace{-0.2cm}

    \subfloat[\label{fig:office}]{
		\includegraphics[width=0.18\textwidth]{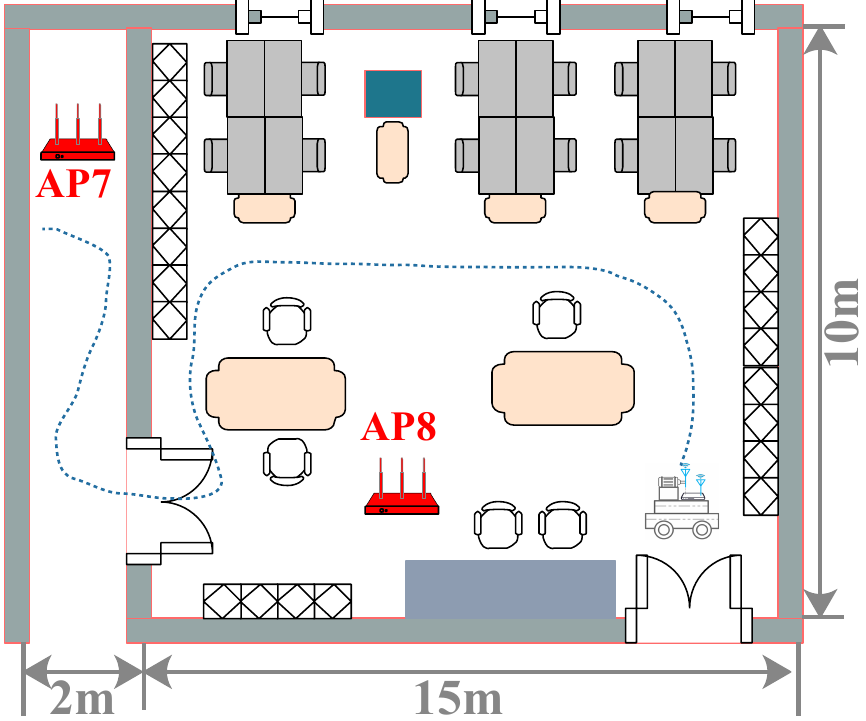}}
	\quad  
 \quad
 \quad
    \subfloat[\label{fig:antenna_layouts}]{
	    \includegraphics[width=0.31\linewidth]{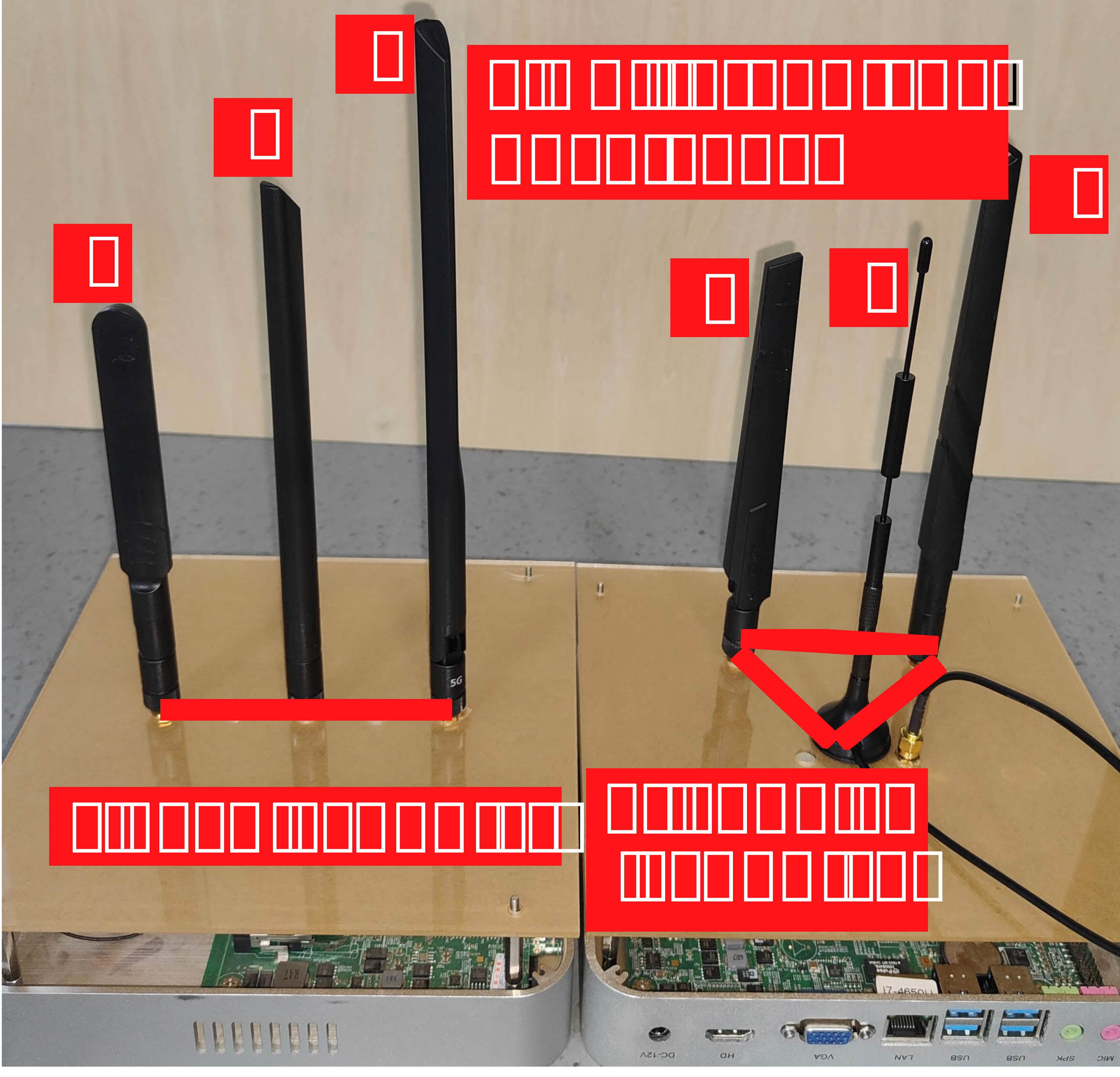}}
     \vspace{-0.2cm}
	\caption{\textbf{Platform implementation and experimental setup:} (a) \oursystem. (b) Effects of bot's position error ($\Delta r \rightarrow \Delta \beta$). (c) Experimental deployment. (d) Hall: A weak-multipath-level scenario. (e) Office: A strong-multipath-level scenario. (f) Six different types of antennas and two different antenna layouts.}
	\label{fig:setup}
	\vspace{-0.35cm}
\end{figure}

\section{Platform Implementation} \label{sec:implementation}
As shown in Fig.~\ref{fig:hardware}, we build \oursystem \ based on an IPC with WiFi NIC IWL5300, and with Intel Core i7-5550U CPU, running on Ubuntu 14.04 LTS. We install~\emph{Linux 802.11 CSI Tool} on the IPC and keep it in communication with the WiFi APs to collect CSI~\cite{Halperin_csitool}. The local antennas are perpendicular polarized omnidirectional dipole antennas with an element length of \SI{4}{cm}. We use a two-phase stepper motor UMot 57HS5417-21-500U~\cite{umotmotor} with a motion control suit and a two-way relay to control the rotation of local antenna 1. We add electromagnetic shielding material~\cite{absorbing} around the local antennas to shield most of the reflected path signal. We fix a high-precision gyroscope WT61C~\cite{gyroscope} at the center of the local antenna 1 to measure this antenna's angles in physical space. We then mount the above system on the TurtleBot platform~\cite{turtlebot}, a low-cost open-source robotics development kit. We mount the Hokuyo UTM-30LN LIDAR~\cite{hokuyo} to capture most obstacles in the environment. We place the local antennas on the top of TurtleBot to have the widest field of view. TurtleBot is controlled via the Robot Operating System (ROS Kinetic), giving us access to a number of software packages for SLAM and navigation. We choose Gmapping~\cite{gmapping} as the SLAM algorithm to construct the map, and make TurtleBot autonomously navigate to several target points.

\subsection{Effects of Bot's Position Error}
We use Gmapping as the SLAM algorithm and navigate the robot, so the target point of the navigation has errors.  We first test the reported position error of the robot, and we can obtain that the median error $\Delta r$ is around \SI{10}{cm}. \oursystem \ requires that the local antenna 1 must be rotated within the plane perpendicular to the LoS path. Thus, the error of robot position is reflected in the mapping of the target antenna to another plane: $\beta_k \rightarrow (\beta_k+ \Delta \beta_k)$, as shown in Fig.~\ref{fig:bot_error}, the error of this A\_angle is $\Delta \beta_k$. When the propagation distance $d > \SI{3}{m}$, we can get $\Delta \beta < \arctan{\frac{0.1}{3}} \approx \SI{1.9}{^\circ}$, so the effects on $\Delta \beta_k$ can be neglected in Equation~\ref{con:normal_vector2}. As for the orientation of the robot, we use the high-precision gyroscope WT61C to correct it, so the orientation error can also be ignored.

%% file: section/evaluation.tex
\section{Evaluation} \label{sec:evaluation}

\subsection{Experimental Setup}
We evaluate \oursystem's performance in real world deployment. To this end, we deploy \oursystem \ in two indoor environments. One scenario is the hall with weak-multipath-level that spans \SI{6500}{sq.ft.} in area. As shown in Fig.~\ref{fig:deployment} and Fig.~\ref{fig:hall}, we deploy APs in six different locations in the hall, covering LoS and NLoS (the obstacles are pillar and board). Another scenario is an office with strong-multipath-level that spans \SI{1800}{sq.ft.} in area. As shown in Fig.~\ref{fig:office}, we deploy APs in two different locations in the office, covering different multipath information. Across these eight different APs, as shown in Fig.~\ref{fig:antenna_layouts}, we cover two antenna geometries, linear and triangular layouts, and cover six antenna types (note that they are all perpendicularly polarized omnidirectional antennas). We use a high-precision gyroscope WT61C as the ground truth for the orientations of the target antennas.

\subsection{Microscopic Benchmark} 
\begin{figure}[t]
    \centering
    \subfloat[\label{fig:baseline_accuracy}]{
	    \includegraphics[width=0.22\textwidth]{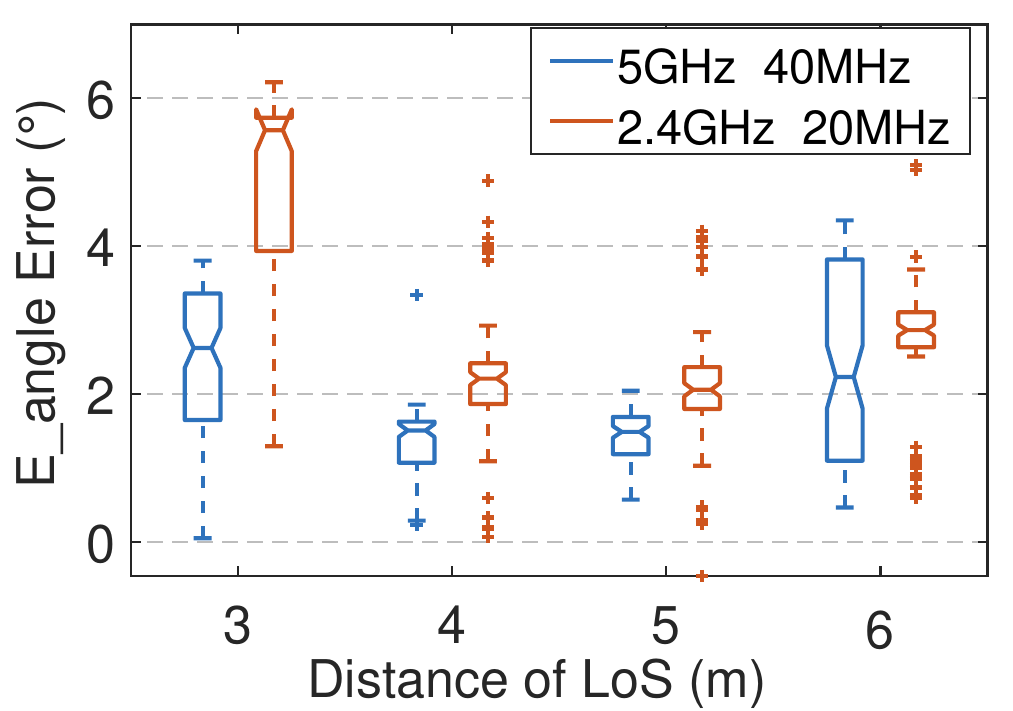}}
	\quad  
	\subfloat[\label{fig:projection_accuracy}]{
	        \includegraphics[width=0.22\textwidth]{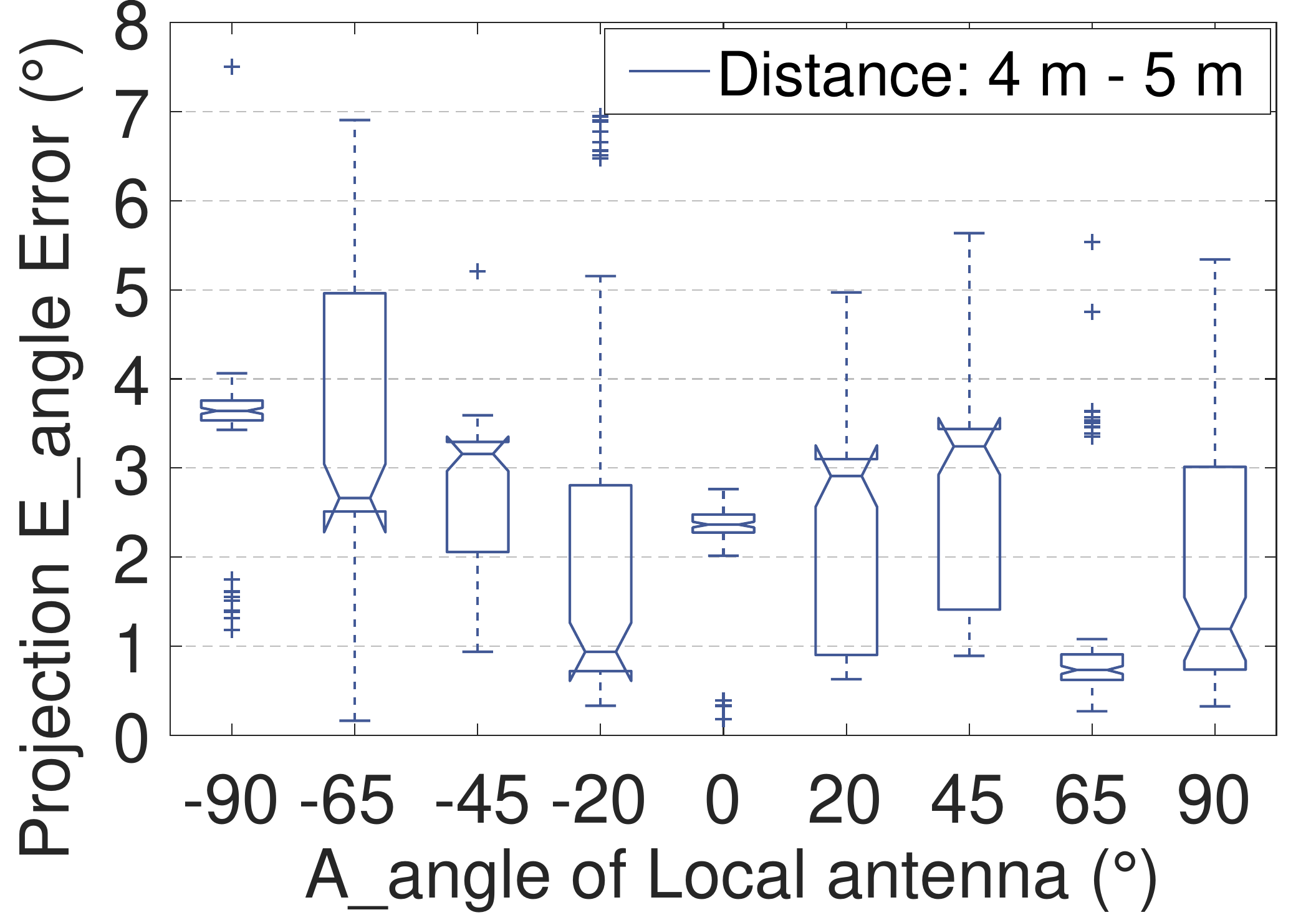}}
    \vspace{-0.18cm}
    \caption{\textbf{Baseline accuracy:} (a) Accuracy of E\_angle for different bands and distances. (b) Accuracy of projection E\_angle for different A\_angle.}
    \label{fig:Microscopic Experiments}
	\vspace{-0.3cm}
\end{figure}

\textbf{Baseline accuracy of E\_angle.}
To verify the baseline accuracy of E\_angle, we fix one target antenna on a stepper motor and rotate it to \SI{0}{^\circ}, \SI{20}{^\circ}, \SI{45}{^\circ}, \SI{70}{^\circ}, \SI{90}{^\circ} (\textit{i.e.}, $\alpha_{obj}$) as depicted in Fig.~\ref{fig:deployment}, and then estimate them. We place robot in LoS condition and rotate the local antenna 1 from \SI{-90}{^\circ} to \SI{90}{^\circ} (\textit{i.e.} $\alpha_{i}$) with a step length of \SI{2}{^\circ} for a total of 90 states. In each state, we collect one second CSI to calculate the received power ratio, and repeat this process 100 times. It is worth noting that the target antenna is parallel to the rotation plane of the local antenna, that is, the A\_angle of the local antenna and the target antenna are always consistent (\textit{i.e.}, $\beta_{obj} = \beta_{k}$). Considering that APs are placed in different locations resulting in different LoS propagation distances. To mimic this, we set the LoS distances to \SI{3}{m}, \SI{4}{m}, \SI{5}{m}, \SI{6}{m} and collect CSI to estimate E\_angle respectively. Further, APs usually operate at \SI{2.4}{GHz} or \SI{5}{GHz} ISM bands with \SI{20}{MHz} or \SI{40}{MHz} bandwidth, so we collect CSI with the same setup for both \SI{5}{GHz}/\SI{40}{MHz} and \SI{2.4}{GHz}/\SI{20}{MHz}. The baseline accuracy of E\_angle is shown in Fig.~\ref{fig:baseline_accuracy}, and the E\_angle errors under all conditions are below \SI{6}{^\circ}. Among them, the accuracy of the LoS distance \SI{4}{m} and \SI{5}{m} is better than that of \SI{3}{m} and \SI{6}{m}. This is due to platform implementation. Specifically, the phase center offset of local antenna and the error of robot are larger when the distance is short, and the multipath effect increases when the distance is long. Hence, in order to obtain better accuracy, we control the LoS distance to the optimal \SI{4}{m}-\SI{5}{m}. In addition, the accuracy of \SI{5}{GHz}/\SI{40}{MHz} is better than that of \SI{2.4}{GHz}/\SI{20}{MHz}, which is caused by the different influence of CSI measurement noise.

\begin{figure}[t]
    \centering
    \subfloat[\label{fig:elevation_accuracy}]{
	    \includegraphics[width=0.22\textwidth]{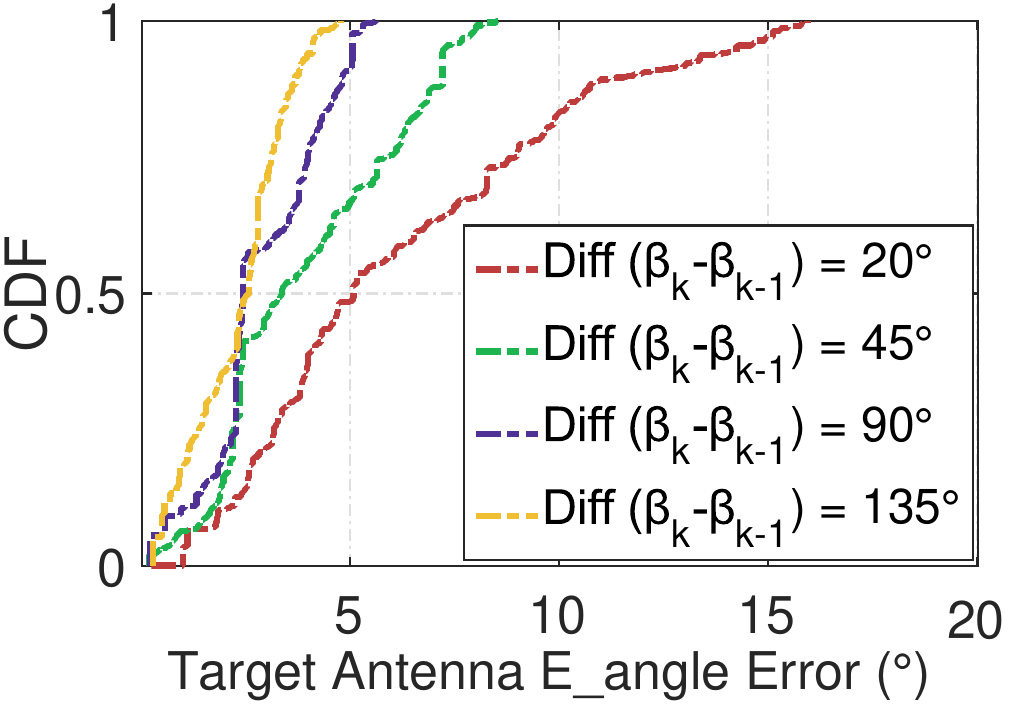}}
	\quad
    \subfloat[\label{fig:azimuth_accuracy}]{
	    \includegraphics[width=0.22\textwidth]{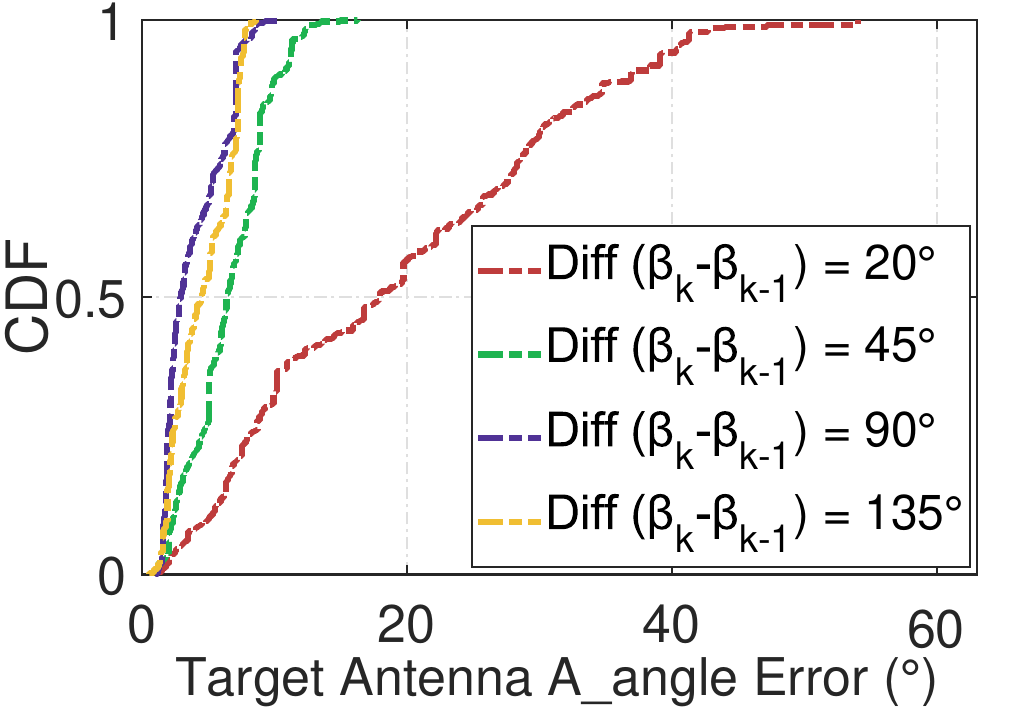}}
    \vspace{-0.18cm}
    \caption{\textbf{Physical orientation accuracy of target antenna:} (a) Estimation accuracy of E\_angle. (b) Estimation accuracy of A\_angle.}
    \label{fig:Microscopic Experiments 2}
	\vspace{-0.3cm}
\end{figure}

\textbf{Effect of A\_angle on projection E\_angle.} 
Since we rotate the local antenna 1 in different vertical planes and obtain the E\_angle of its projection in each vertical plane, we need to verify the accuracy of the projection E\_angle in different vertical planes, \ie, the effect of A\_angle on projection E\_angle. We fix one target antenna on a stepper motor and rotate it to \SI{10}{^\circ}, \SI{30}{^\circ}, \SI{60}{^\circ} (\textit{i.e.} $\alpha_{obj}$) and A\_angle is fixed to \SI{90}{^\circ} (\textit{i.e.} $\beta_{obj}$). Then we collect CSI by rotating the local antenna 1 in the vertical planes with A\_angle of \SI{-90}{^\circ}, \SI{-65}{^\circ}, \SI{-45}{^\circ}, \SI{-20}{^\circ}, \SI{0}{^\circ}, \SI{20}{^\circ}, \SI{45}{^\circ}, \SI{65}{^\circ}, \SI{90}{^\circ} (\textit{i.e.} $\beta_k$) and calculate the projection E\_angle respectively, and we repeat this process 100 times. We use Equ.~\ref{con:inner_product} in combination with $\sin^2{\hat \alpha_{\beta_k}} + \cos^2{\hat \alpha_{\beta_k}} = 1$ to calculate the ground truth of the projection E\_angle. As shown in Fig.~\ref{fig:projection_accuracy}, the projection E\_angle errors are below \SI{7}{^\circ} for different vertical planes.

\textbf{Physical orientation accuracy of target antenna.} 
Based on the above two baseline tests, we estimate the physical orientation of the target antenna in the hall scenario. We set the three antennas on each WiFi AP to different orientations, for example, 
AP2 = $\{(\SI{0}{^\circ}, \SI{0}{^\circ}), (\SI{45}{^\circ}, \SI{0}{^\circ}), (\SI{30}{^\circ}, \SI{-90}{^\circ})\}$. For each AP, we rotate the local antenna 1 with a step length of \SI{2}{^\circ} in the vertical planes around it to collect CSI, where A\_angle of the vertical planes are \SI{-90}{^\circ}, \SI{-65}{^\circ}, \SI{-45}{^\circ}, \SI{-20}{^\circ}, \SI{0}{^\circ}, \SI{20}{^\circ}, \SI{45}{^\circ}, \SI{65}{^\circ}, \SI{90}{^\circ} (\textit{i.e.} $\beta_k$). We then estimate the physical orientations of the target antennas based on some two vertical planes combined with least squares, and the results are shown in Fig.~\ref{fig:elevation_accuracy} and Fig.~\ref{fig:azimuth_accuracy}. We further compare the estimation accuracy using two vertical planes with different spacing (\textit{i.e.} Diff($\beta_k - \beta_{k-1}$)). It can be seen from Fig.~\ref{fig:elevation_accuracy} and Fig.~\ref{fig:azimuth_accuracy} that as the spacing of the vertical planes increases, the estimation error decreases, and the best accuracy is achieved when the spacing is \SI{90}{^\circ}. At this time, the median error of E\_angle is \SI{3}{^\circ}, and the median error of A\_angle is \SI{4}{^\circ}. The reason for this is that when the vertical plane spacing is \SI{90}{^\circ}, the effect between them is minimal. Therefore, we propose to use two vertical planes spaced by \SI{90}{^\circ} to estimate the physical orientation of the antenna. Of course, when some narrow areas cannot meet \SI{90}{^\circ}, the spacing can be reduced.

\subsection{Macroscopic Benchmark}

\begin{figure}[t]
    \centering
    \subfloat[\label{fig:different_types}]{
	    \includegraphics[width=0.52\linewidth]{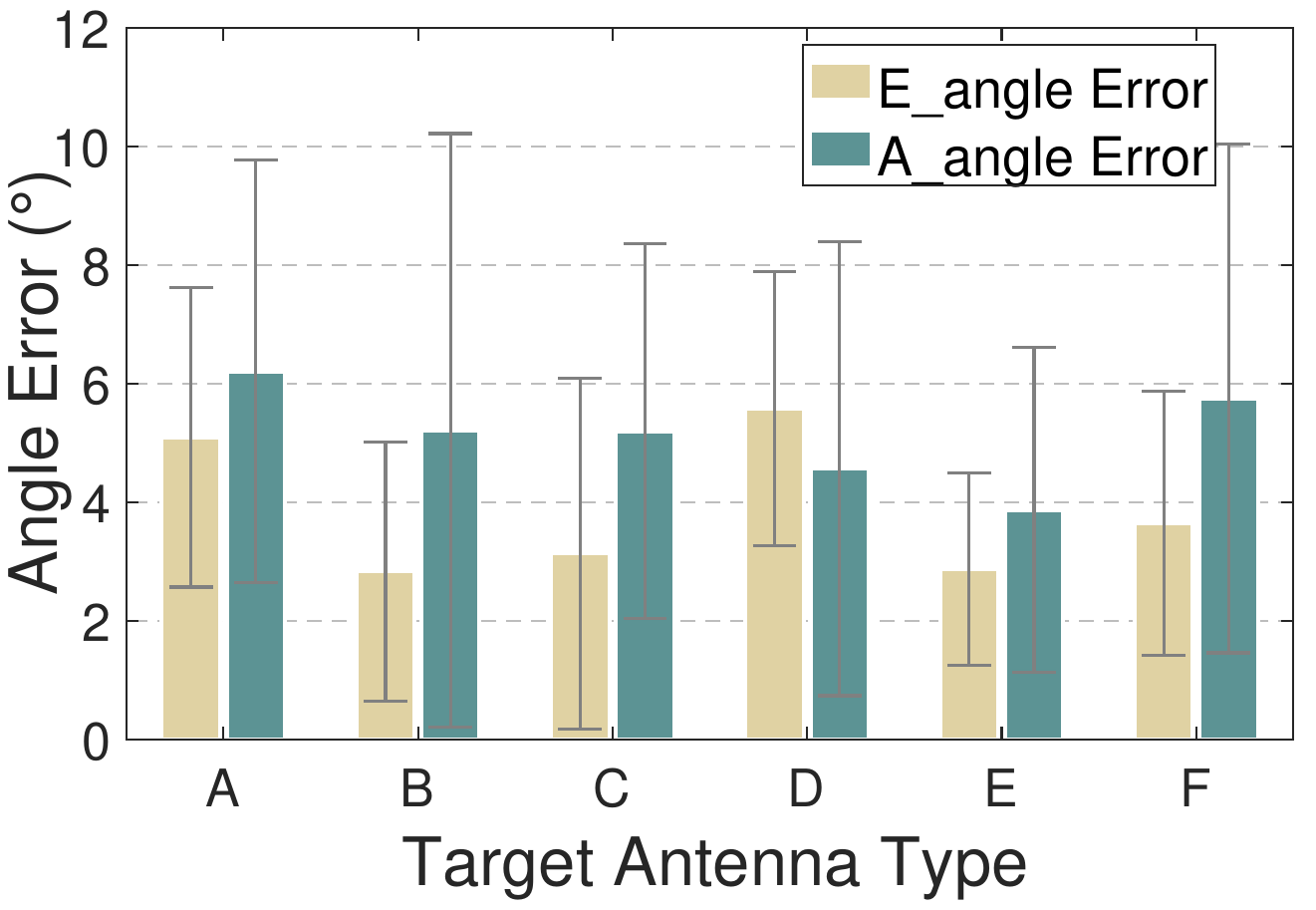}}
	\quad
    \subfloat[\label{fig:different_layouts}]{
	    \includegraphics[width=0.34\linewidth]{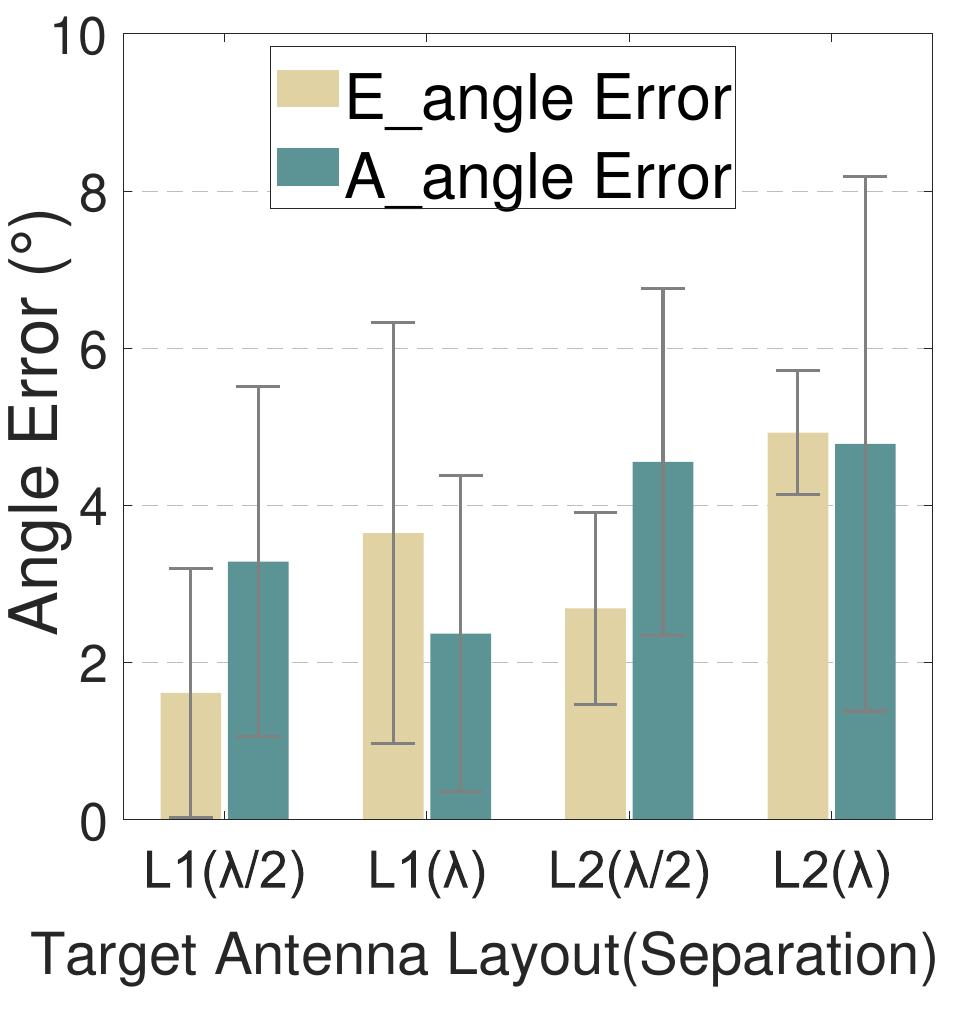}}
    \vspace{-0.18cm}
    \caption{\textbf{Impact of target antenna types and layouts:} (a) Accuracy of angles for six different antenna types. (b) Accuracy of angles for two different antenna layouts (L1: triangular layout, L2: linear layout) and two different antenna separations ($\lambda/2$ and $\lambda$).}\label{fig:target_antenna}
	\vspace{-0.3cm}
\end{figure}

\begin{figure}[t]
    \centering
    \subfloat[\label{fig:multipath_E_err}]{
	    \includegraphics[width=0.22\textwidth]{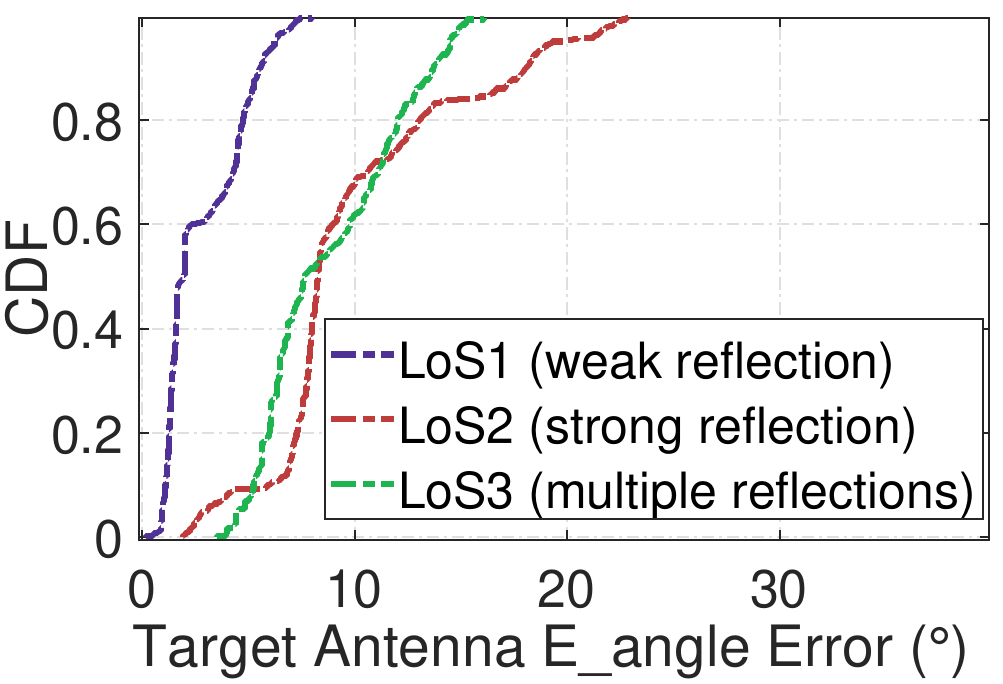}}
	\quad
    \subfloat[\label{fig:multipath_A_err}]{
	    \includegraphics[width=0.22\textwidth]{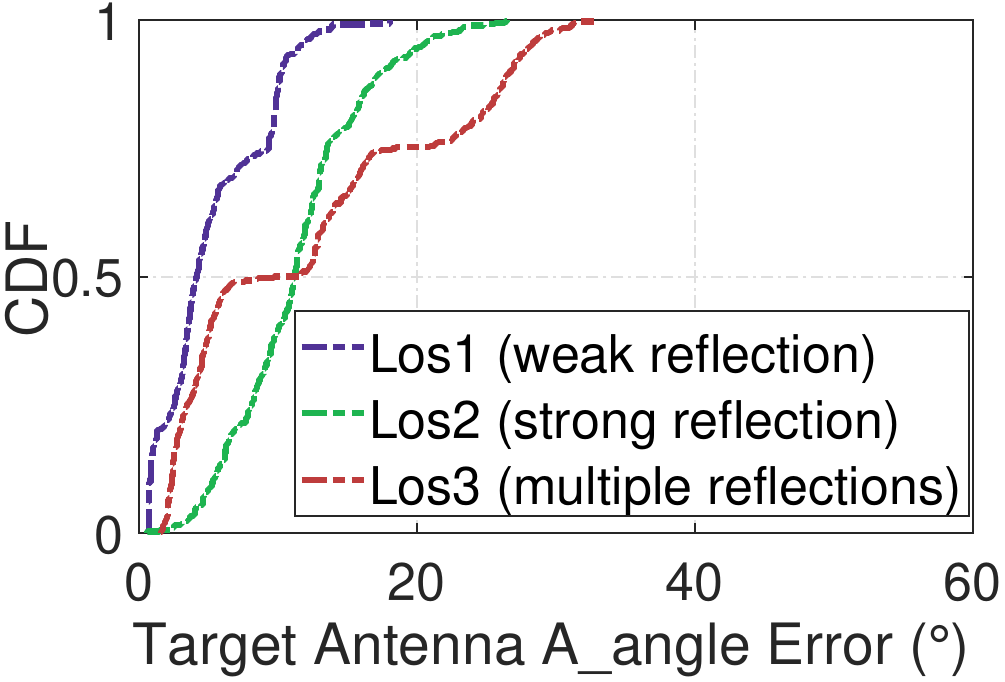}}
     \vspace{-0.35cm}

	\subfloat[\label{fig:nlos_E_err}]{
	    \includegraphics[width=0.22\textwidth]{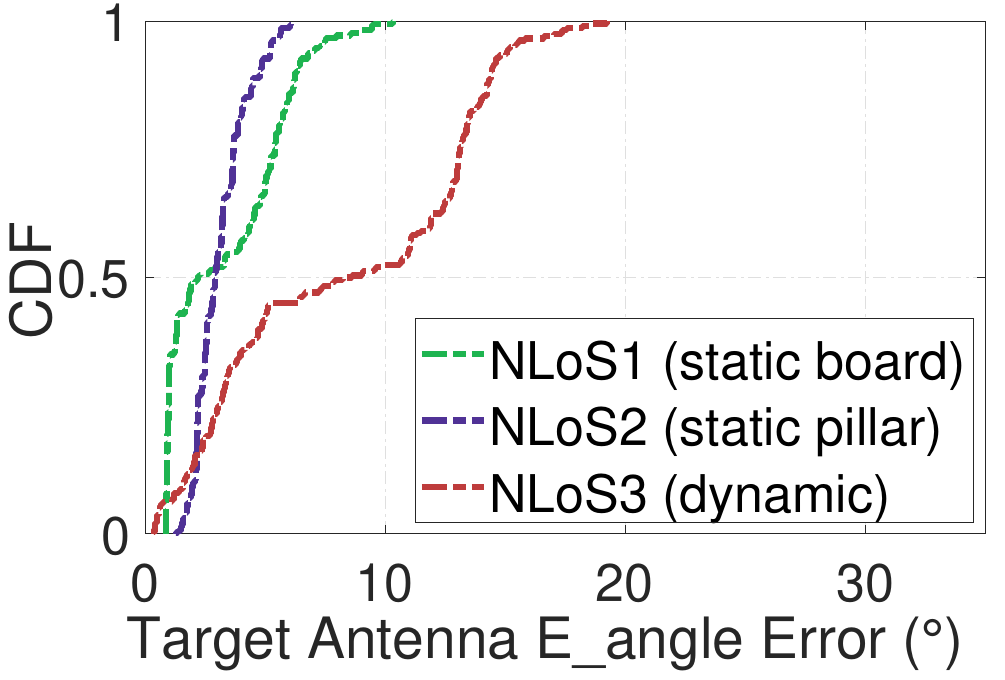}}
	\quad
    \subfloat[\label{fig:nlos_A_err}]{
	    \includegraphics[width=0.22\textwidth]{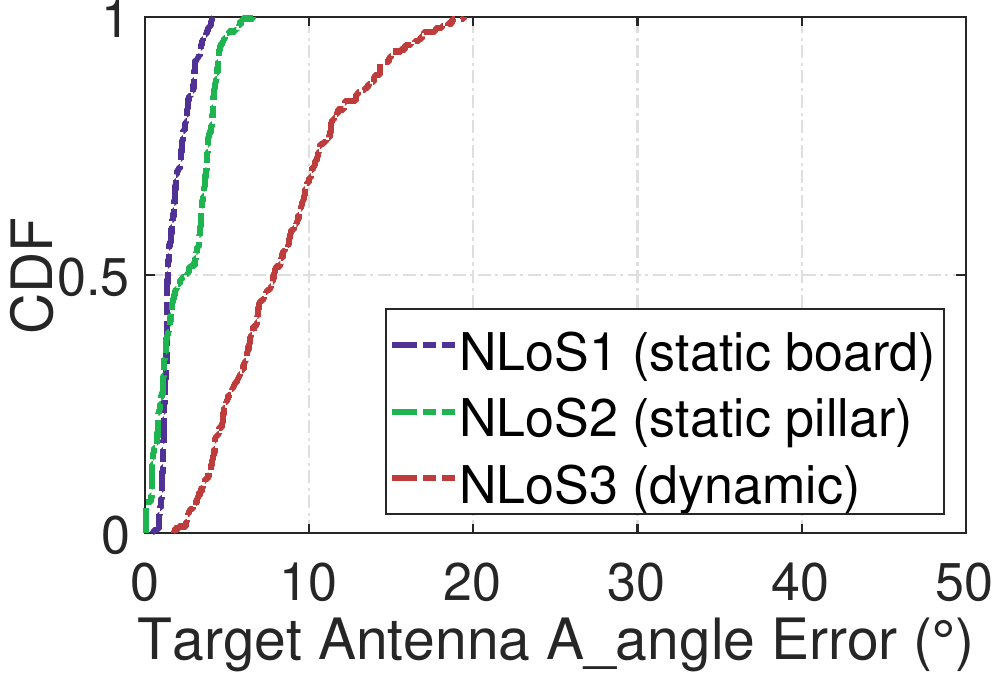}}
    \vspace{-0.18cm}
    \caption{\textbf{Impact of environments:} (a) Accuracy of E\_angle and (b) Accuracy of A\_angle in three different LoS scenarios. (c) Accuracy of E\_angle and (d) Accuracy of A\_angle in three different NLoS scenarios.}\label{fig:different_environment}
    \vspace{-0.3cm}
\end{figure}

\textbf{Impact of antenna types.}
APs may be equipped with different antennas, and different types of antennas have different power lobe patterns, resulting in different power attenuation. We test six different types of antennas in the hall scenario as shown in Fig.~\ref{fig:antenna_layouts}. The heights of these six antennas are different, and the lengths of the antenna elements are also different. The test results are shown in Fig.~\ref{fig:different_types}, it can be seen that for different types of antennas, the median errors of E\_angle are below \SI{6}{^\circ}, and the median errors of A\_angle are also below \SI{6}{^\circ}. In addition, the estimation accuracy of different types of antennas is different because the antenna element itself is slightly bent or tilted.

\textbf{Impact of antenna layouts.}
There are two common geometric layouts of APs, linear and triangular, as shown in Fig.~\ref{fig:antenna_layouts}, and the spacing between antennas is different. We test these two antenna layouts in the hall scenario using one type of antenna, and the antenna spacing is set to $\lambda/2$ and $\lambda$ respectively, a total of four combinations. As shown in Fig.~\ref{fig:different_layouts}, the median errors of both E\_angle and A\_angle are below \SI{6}{^\circ}. Among them, L1 is triangular layout, L2 is linear layout. The reason for accuracy of triangular layouts better than that of linear layouts is that \oursystem \ uses the center of the antenna array as the reference point to construct LoS paths, and the deviation of triangular layouts is smaller than that of linear layouts, which can also be seen from the increase in the estimation error as the antenna spacing increases.

\textbf{Impact of environments.}
We test \oursystem's ability in different environments. 
First, the condition of AP7 is to have one or two strong reflections, and the condition of AP8 is to have multiple reflections. The results are then compared with the weakly reflected conditions in the hall scenario. As shown in Fig.~\ref{fig:multipath_E_err} and Fig.~\ref{fig:multipath_A_err}, the accuracy is reduced in complex multipath, but the median errors are all below \SI{8}{^\circ}. Second, we test in different NLoS, static pillar and wooden, and dynamic NLoS (people moving with objects). As shown in Fig.~\ref{fig:nlos_E_err} and Fig.~\ref{fig:nlos_A_err}, under static NLoS, although there is power attenuation, the attenuation trend is the same, so it has better accuracy. However, under dynamic NLoS, the trend of attenuation is different, so the accuracy decreases, but the median errors are below \SI{8}{^\circ}.

\subsection{Case Studies}
So far, we have studied the performance of \oursystem \ in estimating AP antenna orientation in physical space. 
Then we deploy LocAP and SpotFi, state-of-the-art reverse localization and user localization systems, in the hall scenario to validate the effect of \oursystem. We place four APs at different positions in the hall and make the APs or antennas tilted irregularly. We use a laser rangefinder~\cite{hcjyet} and a WT61C to determine the ground truth of antennas and user.

\begin{figure}[t]
    \centering
	\subfloat[\label{fig:case_study1}]{
	    \includegraphics[width=0.45\linewidth]{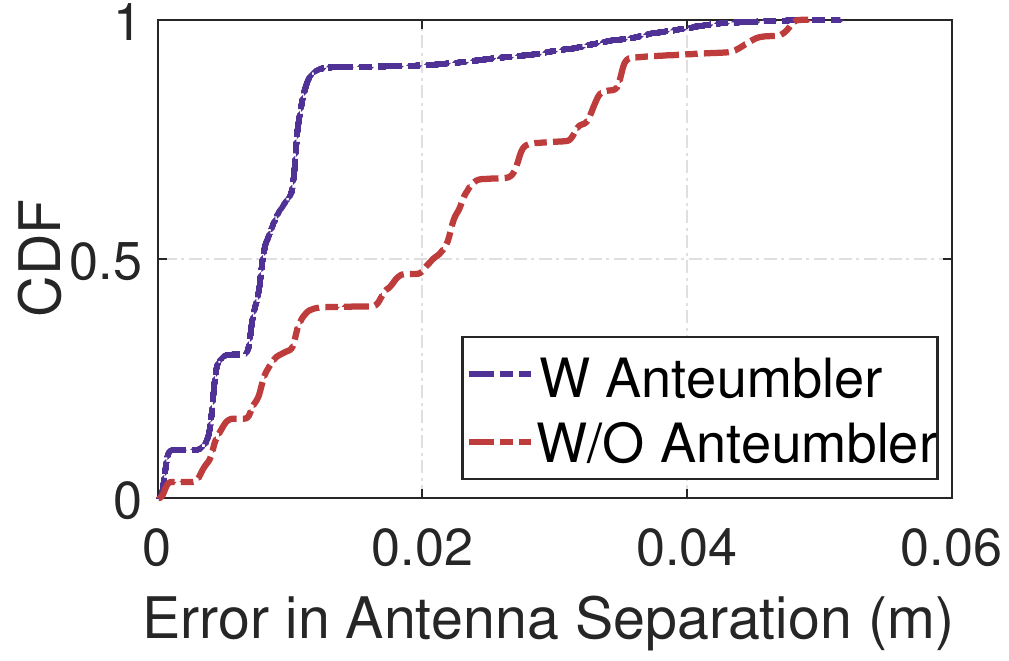}}
	\quad
    \subfloat[\label{fig:case_study2}]{
	    \includegraphics[width=0.45\linewidth]{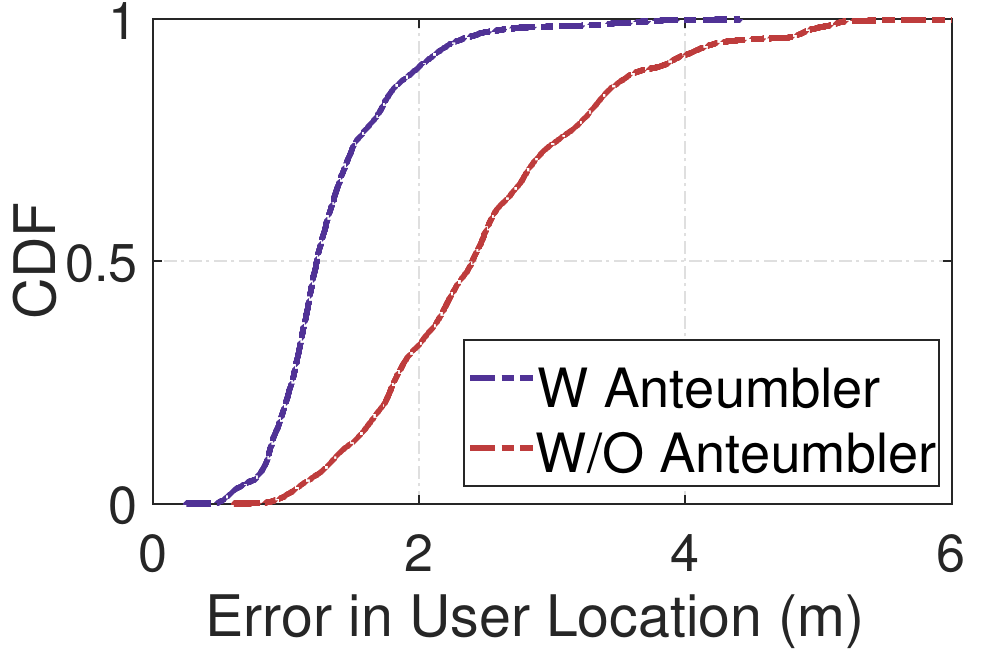}}
    \vspace{-0.18cm}
    \caption{\textbf{Case studies when APs or antennas are tilted:} (a) Antenna separation accuracy in reverse localization with or without \oursystem. (b) User localization accuracy with or without \oursystem.}
    \label{fig:case_study}
	\vspace{-0.3cm}
\end{figure}

\textbf{Case study 1: reverse localization.}
We first deploy LocAP. We set the antenna separation as $\lambda$ and fix the orientation of these APs. Then we move the robot along a straight line, collect one set of CSI every \SI{5}{^\circ}, a total of 20 sets for estimating the antenna separation, and repeat this process 100 times. Next, we deploy \oursystem \ to estimate the physical orientation of these antennas and correct them. After that, the antenna separation is estimated with the same setup. The result is shown in Fig.~\ref{fig:case_study1}. Obviously, when the antennas are tilted irregularly, the error of the antenna separation by LocAP is greatly increased. After the antennas orientations are corrected by \oursystem, the error is reduced by \SI{10}{mm}.

\textbf{Case study 2: user localization.}
Finally, we deploy SpotFi. We set the antenna separation as $\lambda/2$ to obtain the best localization effect. A user with a smartphone connects to these four APs and then moves to 40 different locations in the hall. We collect CSI and use them to calculate user's location. Next, we deploy \oursystem \ to estimate the physical orientation of these antennas and artificially correct them. After that, the user's location is estimated with the same setup. As shown in Fig.~\ref{fig:case_study2}, the error of user location by SpotFi increases greatly when the antennas are tilted irregularly. After the antennas orientations are corrected by \oursystem, the error is reduced by more than \SI{1}{m}.

%% file: section/related_work.tex
\section{Related Work} \label{sec:related}

\textbf{WiFi-based localization.}
In the past two decades, WiFi has been extensively studied for indoor localization and tracking, with two many approaches:~\emph{received signal strength index} (RSSI)-based~\cite{832252,10.1145/1859995.1860016,zhu2013rssi} and CSI-based~\cite{10.1145/2639108.2639139,10.1145/2639108.2639142,10.1145/3210240.3210347,10.5555/2930611.2930623,ArrayTrack,spotfi,10.1145/2789168.2790125}. Many systems achieve decimeter-level localization accuracy using commercial WiFi. However, these localization systems do not take into account the effects of APs' or antennas' orientation errors.
Compared with the above works, we quantitatively analyze the effect of APs' or antennas' orientation errors on localization accuracy, and construct an antenna orientation estimation method that can accurately estimate the orientation of each antenna on the AP. The accuracy of the localization system is guaranteed by correction of the antenna orientation.


\textbf{Reverse localization of WiFi AP.}
In the real world, these AP attributes are often unknown or inaccurate, making WiFi localization difficult to deploy~\cite{locap2020}. Reverse localization is designed to solve this problem by obtaining AP attribute information. There are some works on the reverse localization of WiFi AP~\cite{10.1145/1859995.1860016,8928047,locap2020,MapFi2021}. They can accurately estimate AP locations and antenna spacing. As mentioned in Section~\ref{sec:introduction}, the orientations of APs or antennas are more important. However, they can only measure different orientations of APs in the horizontal plane, and they all assume that all antennas on the AP are parallel. In contrast to these works, \oursystem \ estimates the orientation of each antenna, which in turn ensures the accuracy of WiFi AP reverse localization.


\textbf{Estimation of antenna orientation or tilt angle.}
Traditionally, there are some methods to estimate the elevation and azimuth angles of the antenna. For example, the antenna orientation is manually measured or calibrated using a compass~\cite{compass} or inclinometer~\cite{inclinometer}. There are also integrated systems on the antenna that can measure the antenna orientation~\cite{cpi-2.4aebp,ngabo20183d}. But these are labor intensive or require the antenna to be equipped with sensors. In addition, vision-based method requires sufficient lighting for the antennas to be observed~\cite{yang2022novel}. In this paper, we non-invasively estimate antennas' orientations (elevation and azimuth angles) based on commercial WiFi signals.

%% file: section/conclusion.tex
\section{Conclusion} \label{sec:conclusion}
This paper presents the design and implementation of \oursystem, as far as we are aware, the first attempt to measure the orientation of each antenna of AP in physical space based on WiFi signals. \oursystem \ makes two key technical contributions. First, it includes a spatial angle model, which can estimate the antenna orientation using only the CSI exposed by WiFi chips without extra requirements of the APs. Second, it contains an optimization model that fuses the orthogonality of electric field components, iterative  algorithm and space geometry principles to achieve the accurate measurement of the AP's antennas orientations at the minute-level time cost, and eliminate the influence of propagation distance. Our real-world experiments with different antenna types, different antenna layouts, different AP heights and different environments show that \oursystem \ achieves median errors below \SI{8}{^\circ} for both elevation and azimuth angles. 
Based on accurate measurement of the orientations of all AP antennas, it is expected to help many WiFi-based localization and other sensing systems maintain high accuracy for a long time in the real-world.

%% file: acknow.tex
\section{acknowledgment}
The research is partially supported by the "Pioneer" and "Leading Goose" R\&D Program of Zhejiang (Grant  No. 2023C01029), NSFC with No. 62072424, U20A20181, Key Research Program of Frontier Sciences, CAS. No. ZDBS-LY-JSC001, Anhui Provincial Natural Science Foundation with No. 2308085MF221, Hefei Municipal Natural Science Foundation with No. 2022016, the Fundamental Research Funds for the Central Universities with No. WK3500000008.
